\documentclass[5p,preprint]{elsarticle}

\usepackage{lipsum}

\makeatletter
\newtoks\@eadauthorshort
\def\@author#1#2{\g@addto@macro\elsauthors{\normalsize%
		\def\baselinestretch{1}%
		\upshape\authorsep#1\unskip\textsuperscript{%
			\ifx\@fnmark\@empty\else\unskip\sep\@fnmark\let\sep=,\fi
			\ifx\@corref\@empty\else\unskip\sep\@corref\let\sep=,\fi
		}%
		\def\authorsep{\unskip,\space}%
		\global\let\@fnmark\@empty
		\global\let\sep\@empty}%
	\@eadauthor={#1}
	\@eadauthorshort={#2}
}
\def\@@author[#1]#2#3{\g@addto@macro\elsauthors{%
		\def\baselinestretch{1}%
		\authorsep#2\unskip\textsuperscript{
			\@for\@@affmark:=#1\do{%
				\edef\affnum{\@ifundefined{X@\@@affmark}{1}{\elsRef{\@@affmark}}}%
				\unskip\sep\affnum\let\sep=,}%
			\ifx\@fnmark\@empty\else\unskip\sep\@fnmark\let\sep=,\fi
			\ifx\@corref\@empty\else\unskip\sep\@corref\let\sep=,\fi
		}%
		\def\authorsep{\unskip,\space}%
		\global\let\sep\@empty\global\let\@corref\@empty
		\global\let\@fnmark\@empty}%
	\@eadauthor={#2}%
	\@eadauthorshort={#3}%
}
\gdef\@ead#1{\bgroup\def\_{\string\underscorechar\space}%
	\def\{{\string\lbracechar\space}%
	\def~{\hashchar\space}%
	\def\}{\string\rbracechar\space}%
	\edef\tmpA{\the\@eadauthor}
	\edef\tmpB{\the\@eadauthorshort}
	\immediate\write\@auxout{\string\emailauthor
		{#1}{\expandafter\strip@prefix\meaning\tmpA}{\expandafter\strip@prefix\meaning\tmpB}}%
	\egroup
}
\gdef\emailauthor#1#2#3{\stepcounter{ead}%
	\g@addto@macro\@elseads{\raggedright%
		\let\corref\@gobble
		\eadsep\texttt{#1} (\ifemailshortauthor #3\else#2\fi)\def\eadsep{\unskip,\space}}%
}
\newif\ifemailshortauthor
\makeatother

\usepackage{gensymb}

\usepackage[hyphens]{url}
\usepackage{hyperref}  
\usepackage{graphicx}
\usepackage{caption}
\usepackage{subcaption}
\usepackage[overload]{textcase}
\usepackage{tikz}
\usepackage{rotating}
\usepackage{makecell}
\usetikzlibrary{positioning}

\emailshortauthortrue
\begin{document}

\begin{frontmatter}
		\title{Predicting physical properties of alkanes with neural networks}
		\author[1]{Pavao Santak\corref{correspondingauthor}}{P Santak}
		\cortext[correspondingauthor]{Corresponding author}
		\ead{ps727@cam.ac.uk}
		\author[1]{Gareth Conduit}
		\ead{}
		\address[1]{Theory of Condensed Matter, Department of Physics, University of Cambridge, J.J.Thomson Avenue, Cambridge, CB3 0HE, United Kingdom}
		
\begin{abstract}
We train artificial neural networks to predict the physical properties of linear, single branched, and double branched alkanes. These neural networks can be trained from fragmented data, which enables us to use physical property information as inputs and exploit property-property correlations to improve the quality of our predictions. We characterize every alkane uniquely using a set of five chemical descriptors. We establish correlations between branching and the boiling point, heat capacity, and vapor pressure as a function of temperature. We establish how the symmetry affects the melting point and identify erroneous data entries in the flash point of linear alkanes. Finally, we exploit the temperature and pressure dependence of shear viscosity and density in order to model the kinematic viscosity of linear alkanes. The accuracy of the neural network models compares favorably to the accuracy of several physico-chemical/thermodynamic methods.
\end{abstract}
		
\begin{keyword}
Fragmented data, Neural network, Lubricant, Alkane, Flash point
\end{keyword}		
\end{frontmatter}

\section{Background}
Lubricants are an important component in modern industry. They are used to reduce friction between surfaces, protect them from wear, transfer heat, remove dirt, and prevent surface corrosion to ensure the smooth functioning of mechanical devices. The demand for lubricants makes them an important economic component in oil and gas business, while their importance is only expected to grow. Even as we move towards a future in which fossil fuels will be a less significant source of energy, the lubricant market is expected to grow \footnote{\url{https://www.grandviewresearch.com/press-release/global-lubricants-market}}. 

A typical lubricant product comprises mainly of base oil, which is a mixture of predominantly alkanes that have typically between 18 and 50 carbon atoms. To improve the performance of a base oil, various additives are introduced. Seven physical properties of prime importance for lubricant performance are: melting point, boiling point, flash point, heat capacity, vapor pressure, dynamic viscosity and density. Most individual alkanes with appropriate properties have never been isolated, so relatively little is quantitatively known about their performance. However, data for some alkanes' experimentally determined values is available in TRC Thermodynamic Tables: Hydrocarbons volumes \cite{TRC} or in the DIPPR 801 database \footnote{\url{https://www.aiche.org/dippr/events-products/801-database}}.
	
Lubricants are made from readily available mixtures of predominantly alkanes so it's not certain that current formulations are optimal. Predicting the physical properties of alkanes and understanding the link between alkane structure and lubricant performance would enable the computational design of an optimal base oil, which would motivate the distillation of base oil constituents to approach this optimum in practice.

The physical properties of alkanes that are relevant for base oil lubricant design have previously been modeled with a variety of semi-empirical methods. Wei explored the relationship between rotational entropy and the melting point \cite{MolSim}, while Burch and Whitehead use a combination of molecular structure and topological indices to model the melting point of single branched alkanes with fewer than 20 carbon atoms \cite{MeltingPoint}. To predict the normal boiling point of alkanes, Messerly et al. merged an infinite chain approximation and an empirical equation \cite{Dutch}, while Burch, Wakefield, and Whitehead \cite{Burch} used topological indices and molecular structure to model it for alkanes with fewer than 13 carbon atoms and Constantinou and Gani \cite{Constantinou} developed a novel group contribution method to calculate it for various organic compounds. The semi-empirical Antoine equation is frequently used to model the vapor pressure as a function of temperature.  Mathieu developed a group contribution based method to calculate the flash point of various alkanes \cite{Mathieu}, while Ruzicka and Domalski estimated the heat capacity of various liquid alkanes using a second order group additivity method \cite{Ruzicka}. De La Porte and Kossack have developed a model based on free volume theory to study long chain linear alkane viscosity as a function of temperature and pressure \cite{ViscosityComparison}, Riesco and Vesovic have expanded a hard sphere model to study similar systems \cite{HardSphere}, and Novak has established a corresponding-states model to study viscosity of linear alkanes for the entire fluid region \cite{Lawrence}. 

Purely empirical approaches have also been used in order to predict physical properties of alkanes. For example, Marano et al. develop an empirical set of asymptotic behavior correlations to predict the physical properties of a limited family of alkanes and alkenes \cite{Marano1},\cite{Marano2},\cite{Marano3}. Alqaheem and Riazi, and Needham et al. have explored correlations between different properties \cite{FPBP},\cite{Dayton} to predict the missing values.

While all of these approaches have their own merits, they cannot address the full range of alkanes, as they have a limited range of validity. To accurately predict physical properties for a wide range of alkanes we propose to exploit property-property correlations, molecular structure-property correlations, and semi-empirical equations. Unfortunately, the data set of physical properties of alkanes is fragmented, so to learn the property-property correlations, we need a statistical method that can impute the missing values. One such method is a principle component analysis (PCA) \cite{PCA}, but it delivers accurate results only when variables of interest are linearly correlated. Gaussian processes \cite{GP} is another common approach to handle fragmented data, but they are prohibitively expensive on large datasets and frequently predict large uncertainties for data that is vastly dissimilar to training data, which limits their extrapolative power. 

There is another statistical tool that we could use to predict physical properties of alkanes, artificial neural networks \cite{Bishop},\cite{StatisticalLearning} (ANN) \autoref{fig:NN}. ANN's have undergone rapid development in the last few years, finding applications from image recognition to digital marketing. They have also successfully been used to model physical properties of various organic compounds. For example, Suzuki, Ebert and Sch\"{u}\"{u}rmann used physical properties and indicator variables for functional groups to model viscosity as a function of temperature for 440 organic liquids \cite{ANNSuzuki} Ali implemented a conceptually similar approach to model vapor pressure as a function of temperature for various organic compounds \cite{ANNAli}. Hosseini, Pierantozzi and Moghadasi, on the other hand use pressure, pseudo-critical density, temperature and molecular weight as neural network inputs to model dynamic viscosity of several fatty acids and biodiesel fuels as a function of temperature \cite{ANNHoss}.

Unfortunately, while they are a powerful statistical tool, artificial neural networks previously used to model physico-chemical and thermodynamic properties of organic compounds are not able to handle fragmented data, which limits their applicability to model physical properties of alkanes. However, the neural networks described in Refs. \cite{Gareth}, \cite{GarethMat}, \cite{MatGareth}, \cite{Tom}, \cite{Conduit} can be trained and run with fragmented data, which enables us to exploit property-property correlations even when data is fragmented. This novel neural network formalism has been used to discover two nickel-based alloys for jet engines \cite{MatGareth}, and two molybdenum alloys for forging hammers \cite{GarethMat}, as well as for imputing and finding errors in databases, with over a hundred errors discovered in commercial alloy and polymer databases \cite{Gareth}. It has also been applied for imputation of assay bioactivity data \cite{Tom}. These ANN's serve as a holistic prediction tool for the physical properties of alkanes, enabling us to exploit the property-property correlations, impute the missing values, and exploit the correlations between molecular structure and physical properties. 

In \autoref{Materials and methods}, we present theory of these neural networks, describe an algorithm to generate the molecular basis, and outline a statistical scheme to identify the most accurate neural network model. In \autoref{Results}, we apply this formalism to predict the physical properties of linear and branched alkanes: in \autoref{res1}, we predict the boiling point and the heat capacity of light branched alkanes; in \autoref{res2}, we predict the vapor pressure of light branched alkanes as a function of temperature; in \autoref{res3}, we predict the flash point of linear alkanes and identify erroneous experimental entries; in \autoref{res4}, we predict the melting point of light branched alkanes and explore physical effects of symmetry and in \autoref{res5} we predict the kinematic viscosity of linear alkanes by exploiting the temperature and pressure dependence of their dynamic viscosity and density. Finally, we summarize our findings in \autoref{Conclusions}. We compare the accuracy of neural network models to competing physico-chemical/thermodynamic methods that have been used to model the same properties on similar systems. We determine the accuracy of our models through the coefficient of determination ($\textit{R}^2$) and average absolute deviation (AAD). We decide to use ($\textit{R}^2$) due to its invariance under the shift in data and data rescaling, which is a very useful property for problems in which the neural networks are used, while we chose AAD due to its simplicity and interpretability.

\section{Theory} \label{Materials and methods}
\subsection{Molecular basis}\label{MolBas}
\begin{figure}
	\centering
	\begin{tikzpicture}
	\node[blue] at (0,0) {C};
	\node[blue] at (1,0) {C};
	\node[blue] at (2,0) {C};
	\node[blue] at (3,0) {C};
	\node[blue] at (4,0) {C};
	\node[blue] at (5,0) {C};		
	\node[blue] at (1,1) {C};
	\node[blue] at (2,-1) {C};
	\node[blue] at (2,-2) {C};
	
	\draw[-,blue, thick] (0.35,0) -- (0.65,0);
	\draw[-,blue, thick] (1.35,0) -- (1.65,0);
	\draw[-,blue, thick] (2.35,0) -- (2.65,0);
	\draw[-,blue, thick] (3.35,0) -- (3.65,0);
	\draw[-,blue, thick] (4.35,0) -- (4.65,0);
	\draw[-,blue, thick] (1,0.35) -- (1,0.65);
	\draw[-,blue, thick] (2,-0.35) -- (2,-0.65);
	\draw[-,blue, thick] (2,-1.35) -- (2,-1.65);
	
	\draw[->,red, thick] (0.5,-0.7) -- (0.5,-0.1);
	\node[draw] at (0.5,-1) {$2^{\rm{nd}}$ parameter};
	
	\draw[->,red, thick] (0.1,0.5) -- (0.9,0.5);
	\node[draw] at (-1.1,0.5) {$3^{\rm{rd}}$ parameter};
	
	\draw[->,red, thick] (3.5,0.7) -- (3.5,0.1);
	\draw[->,red, thick] (3.5,0.7) -- (2.5,0.1);
	\draw[->,red, thick] (3.5,0.7) -- (4.5,0.1);
	\node[draw] at (3.5,1) {$4^{\rm{th}}$ parameter};
	
	\draw[->,red, thick] (3,-1) -- (2.1,-0.5);
	\draw[->,red, thick] (3,-1) -- (2.1,-1.5);
	\node[draw] at (4.2,-1) {$5^{\rm{th}}$ parameter};

	\end{tikzpicture}
	\caption{The molecular basis of 3-ethyl-2-methylhexane comprises the five parameters (9,1,1,3,2).}
	\label{fig:Example}
\end{figure}
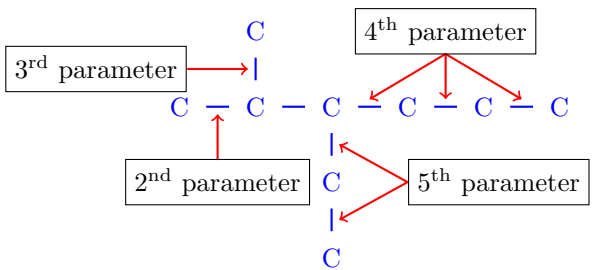

The correlation between molecular structure and physical properties is the backbone of modeling physical properties of alkanes. To exploit these correlations we define a molecular basis that uniquely encodes the structure of every linear, single branched, and double branched alkane into five nonnegative integers. After representing each alkane as a two dimensional graph (\autoref{fig:Example}), these five basis set parameters are:

\begin{enumerate}
	\item The number of carbon atoms.
	\item The smaller number of C-C bonds between the end of the longest carbon chain and its closer branch.
	\item The number of C-C bonds in the branch closer to an end of the longest carbon chain.
	\item The number of C-C bonds between the other end of the longest carbon chain and its closer branch.
	\item The number of C-C bonds in the second branch.
\end{enumerate}

If an alkane has a single branch, the last two basis elements are 0. If an alkane is linear, only the first element is nonzero. This allows the basis set to smoothly pass from straight chain to single to double branched alkane.

\subsection{Neural networks} \label{ANN}
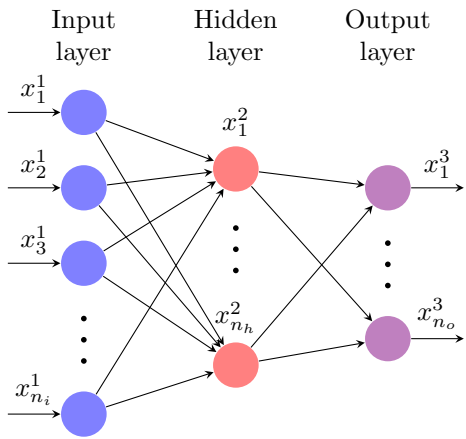
\begin{figure}
		\centering
		\def\layersep{2.5cm}
		
		\tikzset{%
			neuron missing/.style={
				draw=none, 
				scale=2,
				text height=0.333cm,
				execute at begin node=\color{black}$\vdots$			},
		}
		
		\begin{tikzpicture}[x=1.0cm, y=1.0cm, >=stealth]
		
		\foreach \m/\l [count=\y] in {1,2,3}
		{
			\node [circle,fill=blue!50,minimum size=0.6cm] (input-\m) at (0,2.5-\y) {};
		}
		\foreach \m/\l [count=\y] in {4}
		{
			\node [circle,fill=blue!50,minimum size=0.6cm ] (input-\m) at (0,-2.5) {};
		}
		
		\node [neuron missing]  at (0,-1.5) {};

		\foreach \m [count=\y] in {1}
		\node [circle,fill=red!50,minimum size=0.6cm ] (hidden-\m) at (2,0.75) {};
		
		\foreach \m [count=\y] in {2}
		\node [circle,fill=red!50,minimum size=0.6cm ] (hidden-\m) at (2,-1.85) {};
		
		\node [neuron missing]  at (2,-0.3) {};

		\foreach \m [count=\y] in {1}
		\node [circle,fill=violet!50,minimum size=0.6cm ] (output-\m) at (4,1.5-\y) {};
		
		\foreach \m [count=\y] in {2}
		\node [circle,fill=violet!50,minimum size=0.6cm ] (output-\m) at (4,-0.5-\y) {};
		
		\node [neuron missing]  at (4,-0.5) {};
		
		\foreach \l [count=\i] in {1,2,3,n_i}
		\draw [<-] (input-\i) -- ++(-1,0)
		node [above, midway] {$x_{\l}^{1}$};
		
		\foreach \l [count=\i] in {1,n_h}
		\node [above] at (hidden-\i.north) {$x_{\l}^{2}$};
		
		\foreach \l [count=\i] in {1,n_o}
		\draw [->] (output-\i) -- ++(1,0)
		node [above, midway] {$x_{ \l}^{3}$};
		
		\foreach \i in {1,...,4}
		\foreach \j in {1,...,2}
		\draw [->] (input-\i) -- (hidden-\j);
		
		\foreach \i in {1,...,2}
		\foreach \j in {1,...,2}
		\draw [->] (hidden-\i) -- (output-\j);
		
		\foreach \l [count=\x from 0] in {Input, Hidden, Output}
		\node [align=center, above] at (\x*2,2) {\normalsize \l \\  \normalsize layer};

		\end{tikzpicture}
		
		\caption{Schematic of a plain neural network. There are three layers, with input on left through output on the right. Each circle represents a node.}
		\label{fig:NN}
	\end{figure}

Neural networks are a versatile modern statistical tool. They are a universal function approximator \cite{UniApprox} that can recognize patterns that other statistical methods miss. In this section, we describe the theory of the neural networks that we use to predict the physical properties of alkanes. We first present a standard neural network in \autoref{fig:NN}, and then we describe the modifications that enable us to handle fragmented data. 

The standard building block of a neural network is called a node. Each node represents a variable. Nodes are arranged in three types of layers. Every node is denoted by $x^{i}_{j}$, where $x$ is the variable, $i$ is the layer index, and $j$ is the node index. The first layer is called an input layer, comprising the descriptor variables. The second layer is called the hidden layer, and its elements are nonlinear functions of linear combination of input nodes,

\begin{equation}
x^2_{i}=\sigma \bigg(\sum_{j}w^{1}_{ij}x^1_{j}+w^{1}_{0i}\bigg).
\end{equation}

In the above equation, $w^{1}_{ij}$ are called the weights and $\sigma$ is a nonlinear function, commonly known as a transfer function. Our neural networks use $\rm \sigma(x)=tanh(x)$ as the transfer function.The third layer is the output, its elements are linear combinations of nodes in a hidden layer and they represent the estimators for variables of interest,

\begin{equation}
x^3_{i}=\sum_{j}w^{2}_{ij}x^2_{j}+w^{2}_{0i}.
\end{equation}

We train the neural networks by minimizing the cost function 

\begin{equation}
\rm{Cost(W)}=\frac{1}{N}\sum_{i,j}\big(y^{[i]}_{j}-x^{3[i]}_{j}\big)^2.
\end{equation}

\noindent In the above equation, $[i]$ denotes the $i^{\rm{th}}$ example in the training data while ${j}$ denotes the $j^{\rm{th}}$ variable, $y$ denotes the training example, $x^3$ denotes the prediction and N denotes the number of training examples. This form of cost function is called the mean squared error cost function. There are other several other cost functions, such as the mean absolute error cost function, the cross-entropy cost function or the root mean square error cost function \cite{StatisticalLearning}. Minimizing the root mean square error cost function is equivalent to minimization of the mean square error cost function, mean absolute error doesn't have a unique minimum and is used to promote sparsity of the weight matrix, while the cross-entropy cost function is used in classification problems. Modelling physical properties of alkanes is a non-sparse regression problem so the mean square error cost function is an appropriate one to use. Cost function is minimized iteratively by varying the weight matrix $W$ to yield the best model for the training data. To minimize the cost function, we first normalise the data, before we perform a random walk in the weight matrix space until convergence. Some of other commonly used algorithms are the gradient descent and the stochastic gradient descent \cite{StatisticalLearning}, but we have found that the random walk algorithm with a predetermined expected move acceptance probability of 20\% is as accurate and faster than other commonly used algorithms. 

\subsection{Handling sparse data}

\begin{figure}
	\centering
	\includegraphics[width=0.9\linewidth]{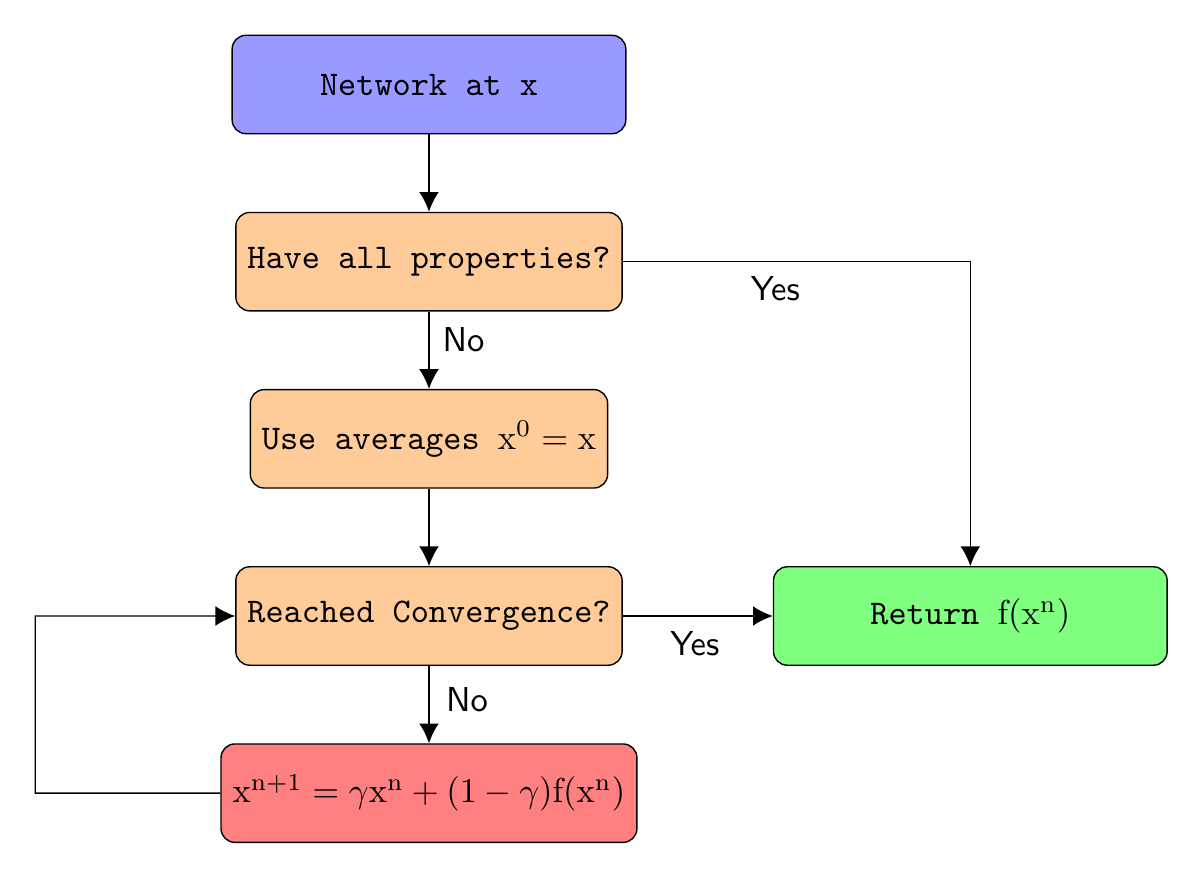}
	\caption{Data imputation algorithm for the vector x. We set $\rm x^0=x$, replacing all missing entries by averages across each assay and, then, iteratively compute $\rm x^{n+1}$ as a function of $\rm x^n$ and $\rm f(x^n)$ until we reach convergence.}
	\label{fig:flowchart}
\end{figure}

The neural networks described above can exploit the correlations between molecular structure and physical properties, but they cannot train from sparse data, as they require all the inputs to give an output. To exploit the property-property correlations we use physical properties as both inputs and outputs of the neural network model, which requires two changes to its architecture. Firstly, during neural network training, the weights $w^2_{ii}$ are set to zero to ensure property predictions are independent of the original value. Secondly, after the network is trained, we replace the missing values with the mean property values and recursively apply the following equation:

\begin{equation}
x^{[n+1]}=\gamma x^{[n]}+(1-\gamma)f(x^{[n]}),
\end{equation} 

where $n$ denotes the iteration step, $f(x)$ is a prediction for $x$ obtained from the neural network, and $\gamma \in [0,1]$ is a mixing parameter. In this manuscript, we use $\gamma=\frac{1}{2}$. We apply the above equation until convergence, before we apply function $f$ again. A schematic of the data imputation algorithm is shown in \autoref{fig:flowchart}. To predict the mean and the uncertainty in the physical property of interest, we train and run six neural networks in parallel, assigning random weights to each data entry for each neural network model. For each neural network $k$, the cost function then takes the following form:

\begin{equation}
\rm{Cost(W)}_{k}=\frac{1}{N}\sum_{i,j}q_{k,i}\big(y^{[i]}_{j}-x^{3[i]}_{j}\big)^2,
\end{equation}
	
\noindent where $\sum_{i}q_{i}=1$.

\subsection{Cross validation} \label{CV}
Training error, which we measure through the coefficient of determination ($\textit{R}^2$) and average absolute deviation (AAD), is a poor indicator of neural network's predictive power, as it underestimates the true error in neural network models. To obtain a better estimate of the neural network model accuracy, we perform cross-validation by splitting the full data set into the training set and the validation set. We use a scheme called leave-one-out cross-validation, in which we train the neural network on all but one data entry in a dataset before we test it against the remaining entry. We repeat this process until neural network has been tested against every entry in a dataset.

\begin{figure}
	\centering
	\includegraphics[width=0.7\linewidth]{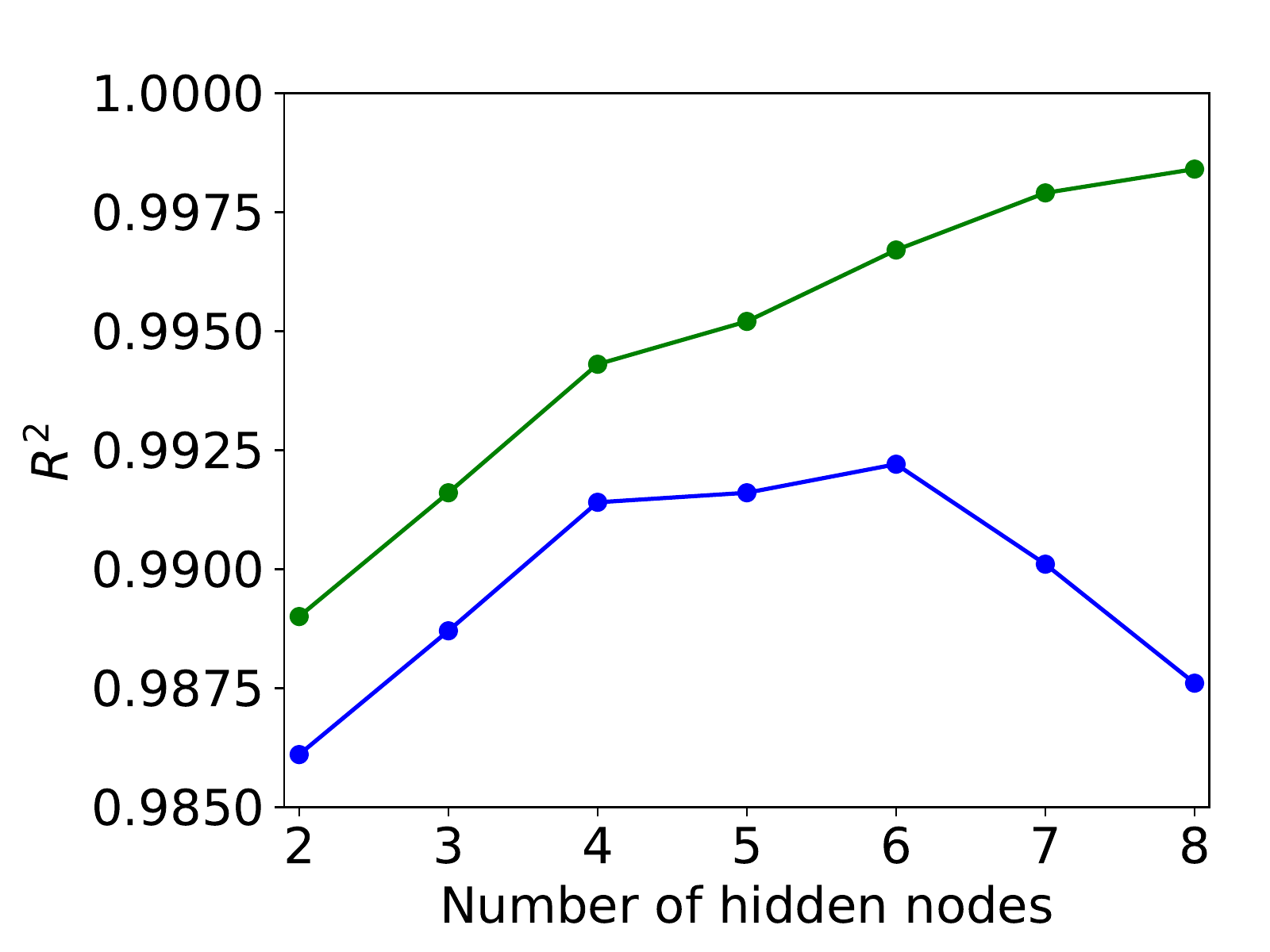}
	\caption{Determining the optimal number of hidden nodes for the boiling point. Green curve represents training $\textit{R}^2$, while the blue curve represents the cross-validation $\textit{R}^2$.}
	\label{fig:HiddenNodes}
\end{figure}

We also use the leave-one-out cross-validation to determine the optimal number of hidden nodes for our neural network. We train neural networks with different number of hidden nodes, perform cross-validation for each of them, and choose an architecture that has the smallest cross validation error. We illustrate this procedure in \autoref{fig:HiddenNodes} by determining the optimal neural network architecture for predicting boiling point of straight, single-branched and double-branched alkanes. In this case, the training error $\textit{R}^2$ increases as a function of number of hidden nodes, but the cross-validation error is the smallest for the neural network model with 6 hidden nodes. Too few hidden nodes are unable to properly capture the behavior while too many hidden nodes overfits the training data.

\section{Results and discussion}\label{Results}
In this section, we apply the formalism presented in \autoref{Materials and methods} to predict the physical properties of alkanes. We first predict the boiling point and the heat capacity of branched alkanes. Then we predict the Antoine coefficients to model the vapor pressure as a function of temperature. We also identify erroneous data in the flash point data from the literature, predict the flash point and establish the connection between the number of molecular symmetries and the melting point. Finally, we exploit the temperature and pressure dependence of dynamic viscosity and density and predict the kinematic viscosity of linear alkanes as a function of temperature.

We work with linear alkanes up to tridecane and with branched alkanes with fewer than 13 carbon atoms. Our data set comprises of experimental values obtained from various online sources (\cite{chemeo}, \cite{cdo}, \cite{nat}), as well as experimental values presented in previous research papers (\cite{Assael1991}, \cite{BALED2014108}, \cite{Caudwell2004}, \cite{doi:10.1021/je800417q}, \cite{HERNANDEZGALVAN200751}, \cite{doi:10.1021/je00049a011}, \cite{SANTOS201746}, \cite{unknown}) and the TRC Thermodynamic Tables \cite{TRC}.

\subsection{Boiling point and heat capacity}\label{res1}
\begin{figure}
	\centering
	\includegraphics[width=0.7\linewidth]{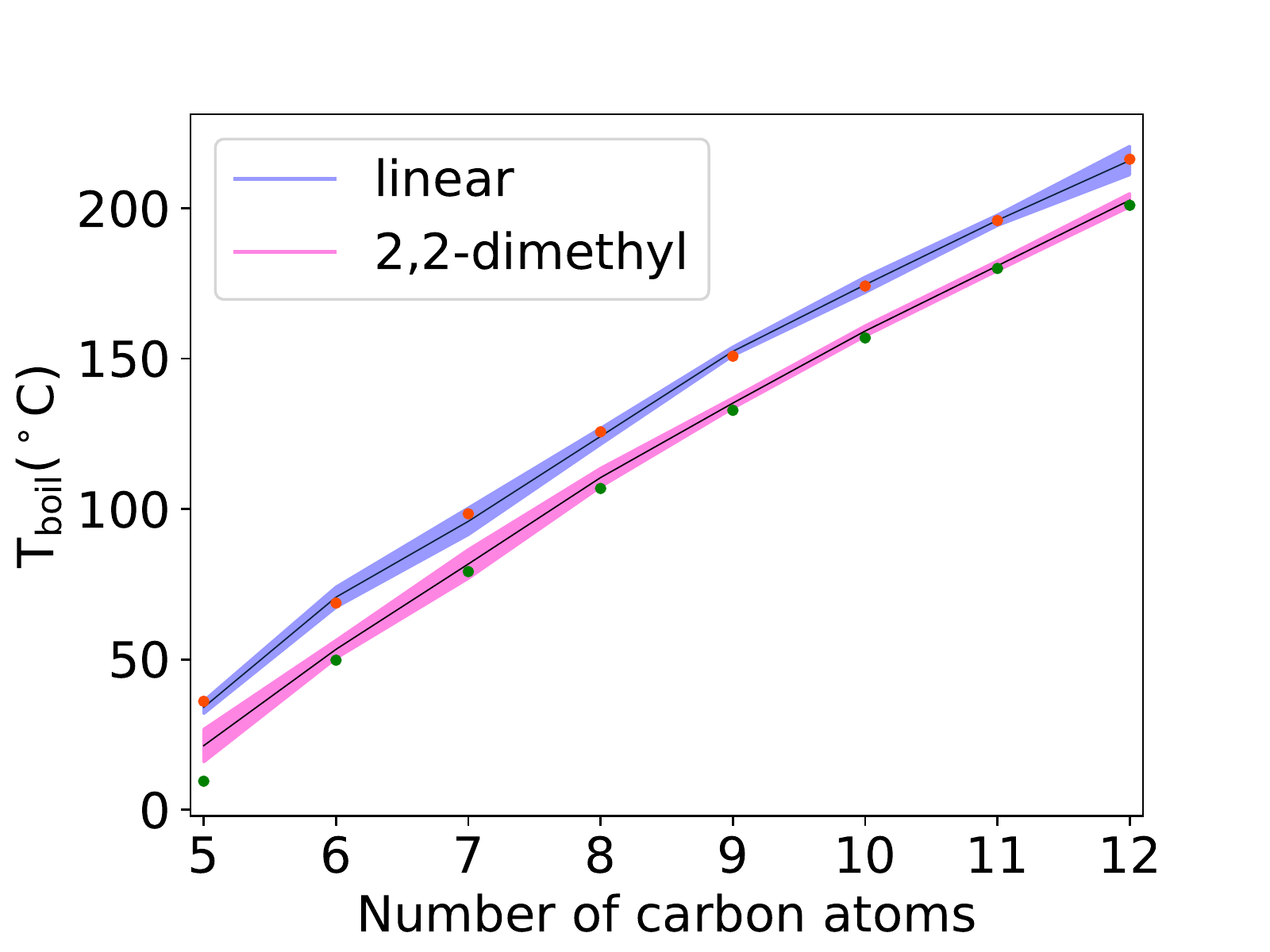}
	\caption{Boiling point vs number of carbon atoms for linear alkanes and the 2,2-dimethyl homologous series. Dots represent the experimental data entries, blue and green lines represent the mean predictions of the neural network while the colored areas represent the uncertainties in the predictions.}
	\label{fig:BP}
\end{figure}

Predicting the boiling point of alkanes is an important step in determining their suitability for use in base oils, as alkanes with higher boiling points stay liquid at higher temperatures. We predict the normal boiling point of branched alkanes with fewer than 13 carbon atoms by training a neural network on an dataset comprised of 188 alkanes \cite{TRC} with molecular basis as the input nodes and 6 hidden nodes (\autoref{CV}), obtaining a cross-validation $\textit{R}^2=0.992$ and an AAD of 1.74$^{\circ}$C, indicating an excellent fit. The high quality of fit to experimental data can be seen in \autoref{fig:BP}.

\begin{figure}
	\centering
	\includegraphics[width=0.7\linewidth]{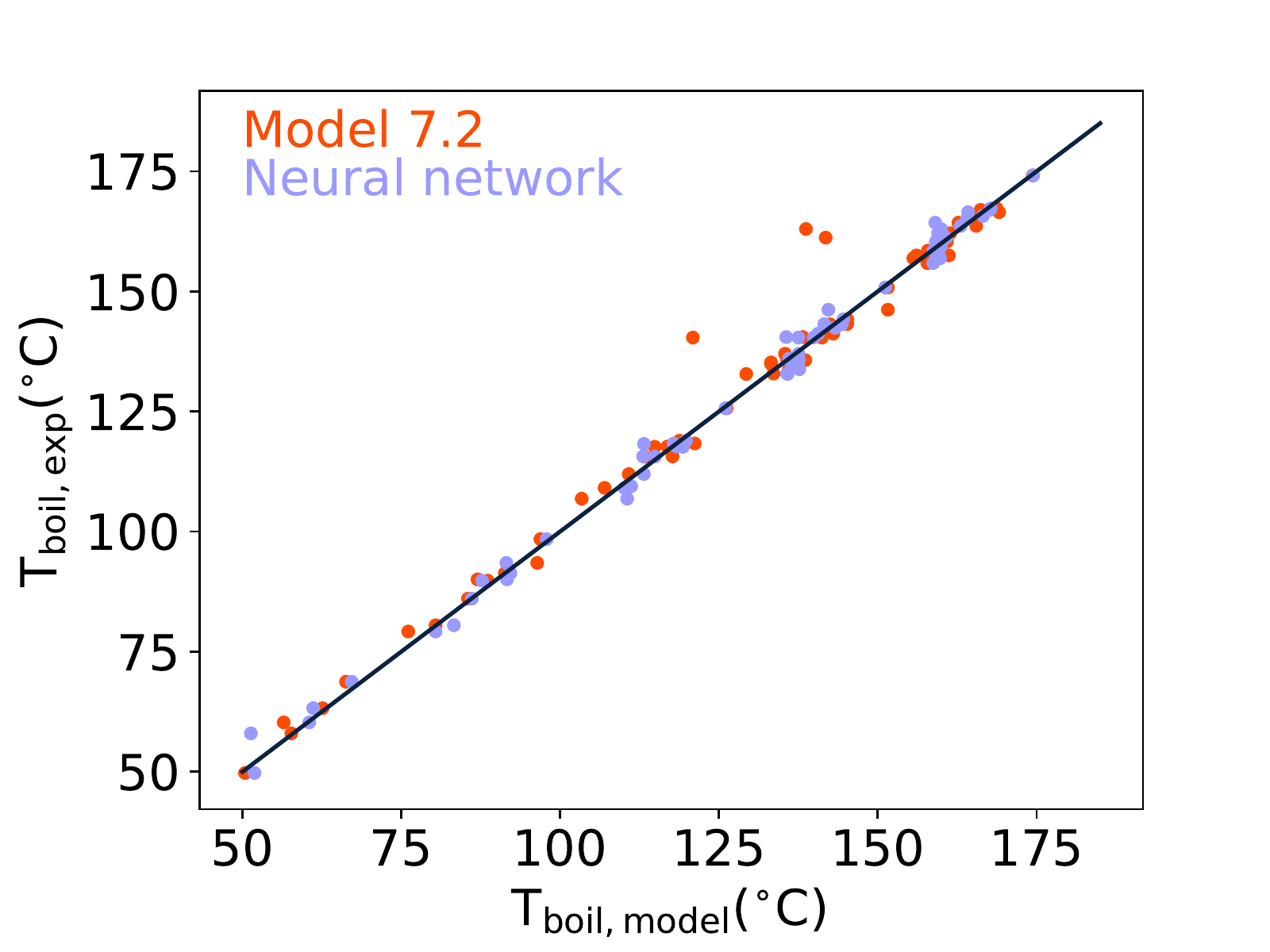}
	\caption{Parity plot for boiling point of alkanes. Our neural network model is compared to the model 7.2 from \cite{Burch}.}
	\label{fig:BPPar}
\end{figure}

After establishing the accuracy of the neural network, we compare our results to two regression models that use molecular structure and topological indices as inputs \cite{Burch}. We compare three models for 62 alkanes whose boiling point all three models predict. Our neural network model outperforms both alternative models (\autoref{tab:table1}). Apart from showing improved accuracy, our neural network model shows greater consistency than two competing models, as the standard deviation in absolute error is 1.43$^{\circ}$C, while the standard deviation in absolute error of model 7.2 is 4.37$^{\circ}$C and 4.75$^{\circ}$C for model 7.3. A parity plot is shown in \autoref{fig:BPPar}. There are several molecules that both models 7.2 and 7.3 mispredict by a significant margin. Absolute deviations for the boiling points of 3-ethyl-2-methyl-heptane, 3-ethyl-3-methyl-pentane and 3-ethyl-3-methyl-heptane and are 19.4$^{\circ}$C, 19.5$^{\circ}$C and 24.3$^{\circ}$C for model 7.2 and similar for model 7.3, while they are 0.44$^{\circ}$C, 0.48$^{\circ}$C and 3.03$^{\circ}$C for the neural network model. 

\begin{figure}
	\centering
	\includegraphics[width=0.7\linewidth]{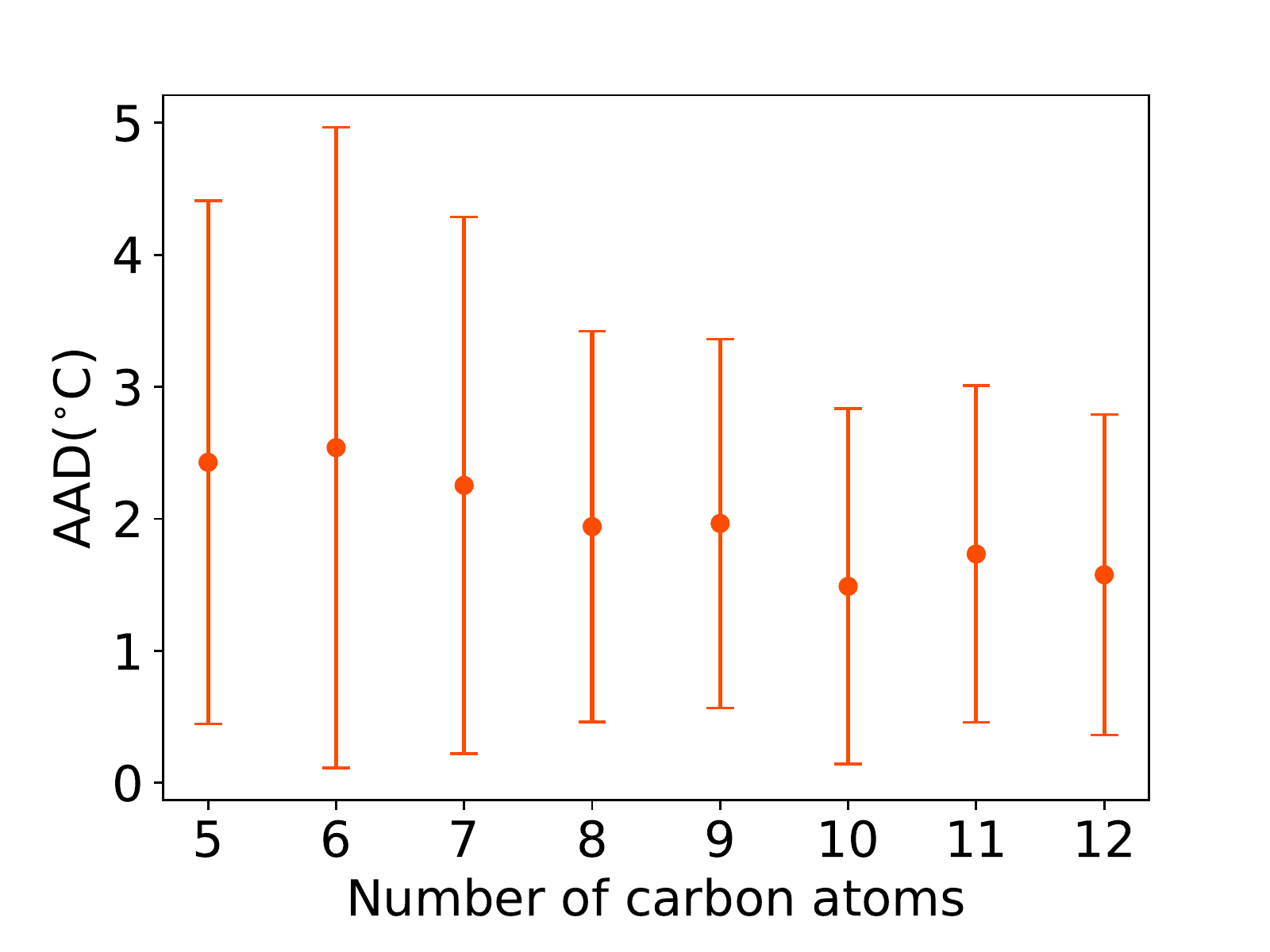}
	\caption{Average absolute deviation of the neural network model for the boiling point as a function of number of carbon atoms.}
	\label{fig:BPAAD}
\end{figure}

Focusing only on alkanes with five or more carbon atoms, we observe that average absolute deviation for structural isomers decreases with increasing molecular weight \autoref{fig:BPAAD}. A decreasing average absolute deviation, as well as greater accuracy and consistency of our predictions compared to other models means that our neural network model can be used to predict the boiling point of alkanes whose boiling point hasn't yet been experimentally measured with higher confidence. 

We observe that adding a branch but keeping the molecular weight constant decreases the boiling point by about 7$^\circ$C. Increasing the length of the branch while keeping molecular weight constant reduces the boiling point by about 2$^\circ$C, while moving the branch by an atom along the longest chain reduces it by about 2$^\circ$C.

\begin{table}[h!]
	\begin{center}
		\begin{tabular}{|c|c|c|} 
			\hline
			\textbf{Method} & \textbf{$\textit{R}^2$} & \textbf{AAD $\rm(^{\circ} C)$}\\
			\hline
			Neural Network  & 0.995 & 1.69 \\
			\hline
			\makecell[c]{Model 7.2 \cite{Burch}} & 0.977 & 2.47\\
			\hline
			\makecell[c]{Model 7.3 \cite{Burch}} & 0.975 & 2.24 \\
			\hline
		\end{tabular}
		\caption{Summary of accuracy of three boiling point models. Our neural network model is compared to two regression models that use molecular structure and topological indices as inputs.}
		\label{tab:table1}
	\end{center}
\end{table} 

We also predict the molar heat capacity of branched alkanes with fewer than 13 carbon atoms at 25$^\circ$C. The larger the molar heat capacity, the more energy an alkane can absorb and transport without a change in temperature, making it more suitable for use in lubricant base oils.

\begin{figure}
	\centering
	\includegraphics[width=0.7\linewidth]{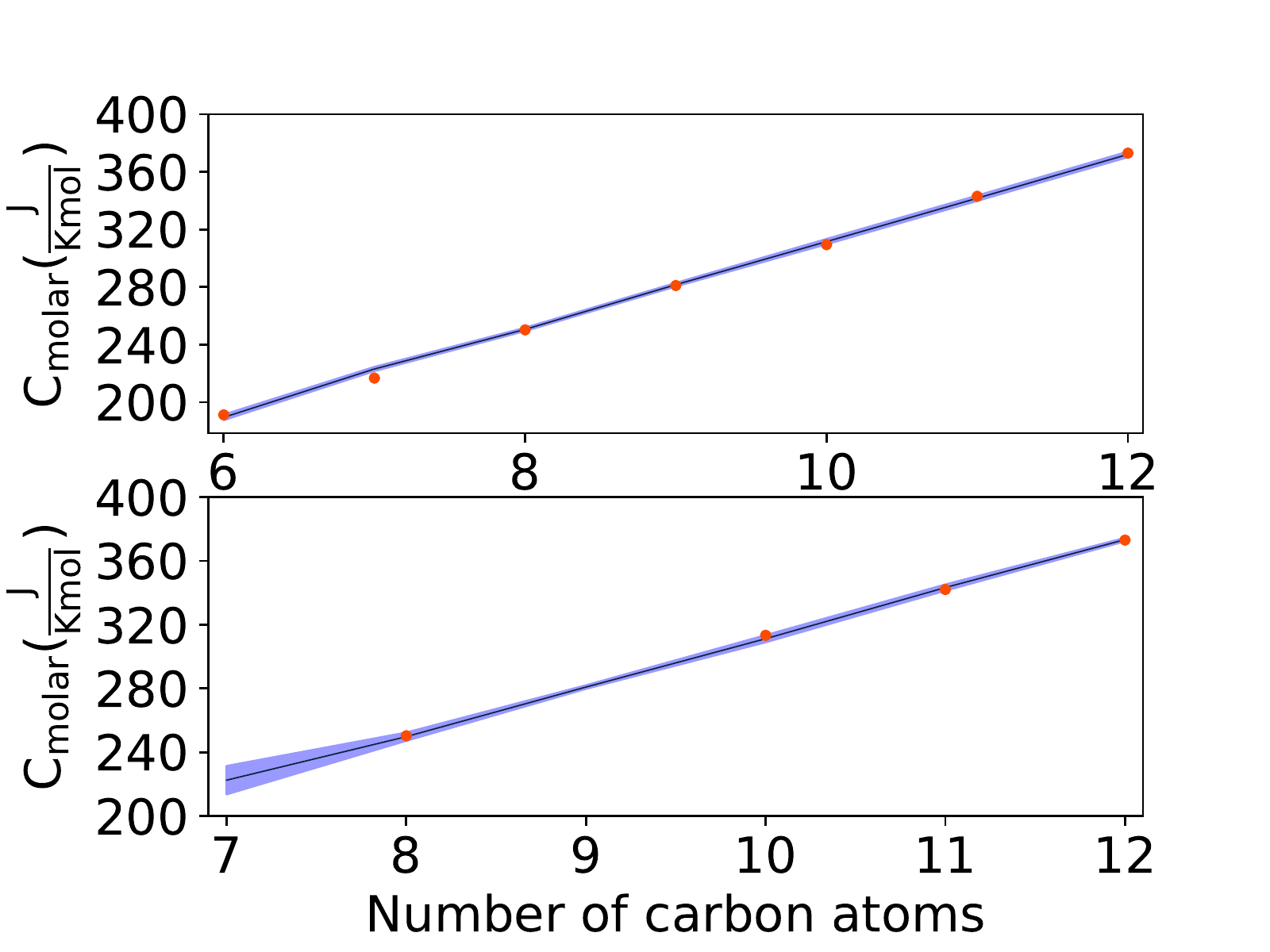}
	\caption{Heat capacity for 3-methyl and 3-ethyl homologous series. In the top figure, we show heat capacity predictions for the 3-methyl series, while in the bottom figure we show predictions for the 3-ethyl series. Orange dots represent experimental data entries, blue line represents mean prediction while the area around it represents the uncertainty.}
	\label{fig:HC}
\end{figure}

After applying the same neural network architecture that we used to predict the boiling point to a dataset comprised of 176 alkanes, we obtain a cross-validation $\textit{R}^2=0.997$, showing an excellent fit. Our dataset doesn't include methane, ethane, propane, butane and 2-methylbutane, as they are not liquids at 25$^\circ$C. We can see the quality of fit for some of our predictions in \autoref{fig:HC}. 

\begin{table}[h!]
	\begin{center}
		\begin{tabular}{|c|c|c|c|} 
			\hline
			\textbf{Method} & \textbf{$\textit{R}^2$} & \textbf{AAD ($\rm{J(molK)^{-1}})$}\\
			\hline
			Neural Network  & 0.996 &  2.10\\
			\hline
			\makecell[c]{Second Order \\ Group Additivity \cite{Ruzicka}} & 0.994 & 2.87\\
			\hline
		\end{tabular}
		\caption{Summary of comparison of accuracies of the neural network model and a second order group additivity model for the heat capacity.}
		\label{tab:table2}
	\end{center}
\end{table}

\begin{figure}
	\centering
	\includegraphics[width=0.7\linewidth]{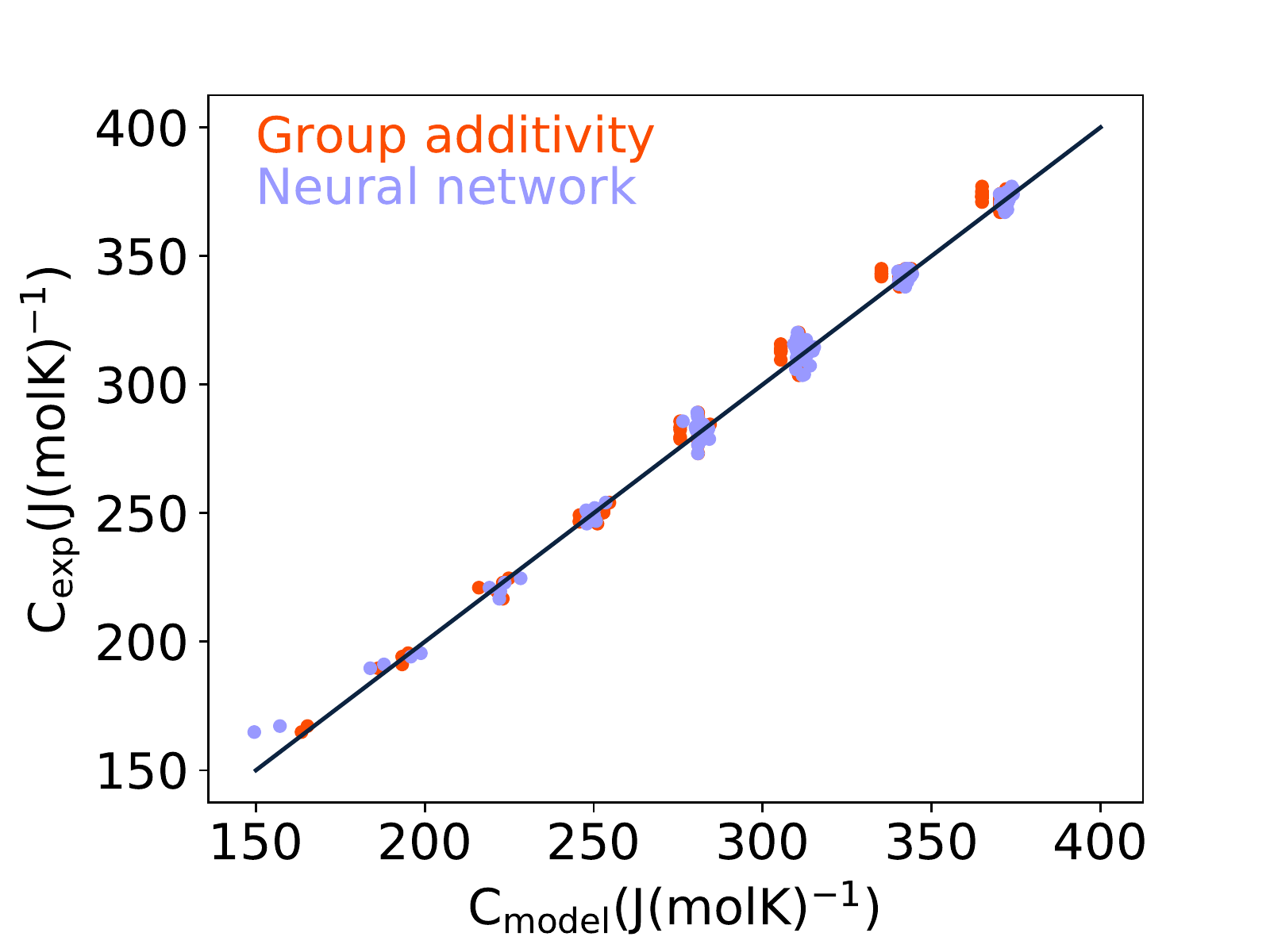}
	\caption{Parity plot for heat capacity of alkanes. Neural network model is compared to the second order group additivity \cite{Constantinou}.}
	\label{fig:HCPar}
\end{figure}

We compare the quality of predictions from the neural network model to those from a model based on second order group additivity \cite{Ruzicka}. The neural network model outperforms the second order group additivity method, giving an AAD of 2.10 $\rm{J(molK)^{-1}}$ (\autoref{tab:table2}). Our model also exhibits greater consistency than the second order group additivity method. Standard deviation in the absolute deviation of our neural network models is 2.04$\rm{J(molK)^{-1}}$, compared to 2.87$\rm{J(molK)^{-1}}$ for the second order group additivity method. We show a parity plot for both models in \autoref{fig:HCPar}.

\begin{figure}
	\centering
	\includegraphics[width=0.7\linewidth]{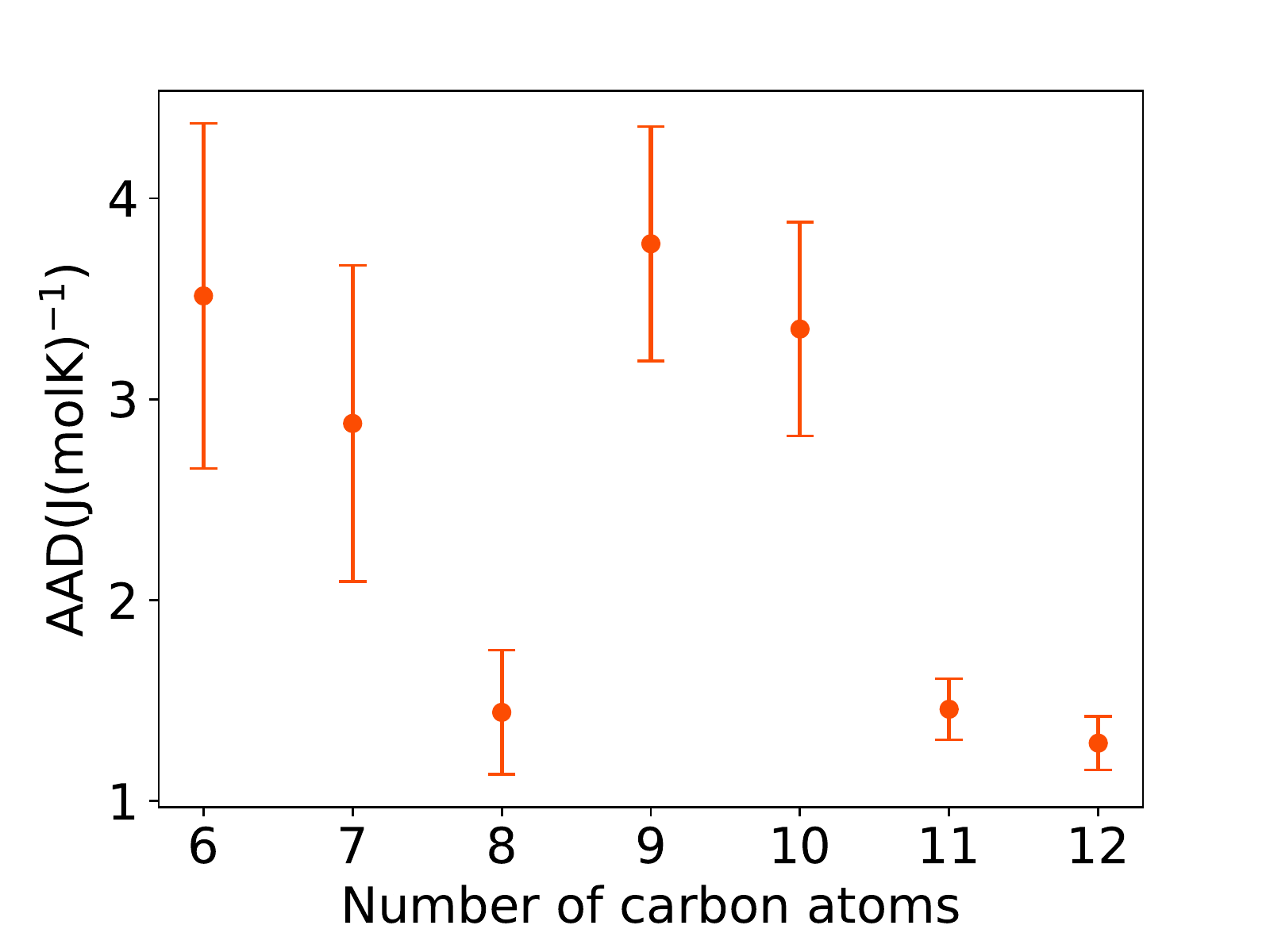}
	\caption{Average absolute deviation of the neural network model for the boiling point as a function of number of carbon atoms.}
	\label{fig:HCAAD}
\end{figure}

We also investigate the accuracy of our models as a function of carbon atoms for all the alkanes with more than 5 and fewer than 13 carbon atoms \autoref{fig:HCAAD}. Unlike for the boiling point, we do not observe decrease of average absolute deviation with increase in molecular weight. While the average absolute deviation is the smallest for the structural isomers of dodecane, it is the largest for the isomers of nonane. Nonetheless, increased accuracy and consistency of our model result in higher confidence in using neural networks to predict molar heat capacity of alkanes whose heat capacity is unknown. Our results indicate that the molar heat capacity is approximately an increasing linear function of number of carbon atoms, while the effects of adding a branch, increasing its length or moving it along the longest carbon chain are negligible.

\subsection{Vapor pressure}\label{res2}
Vapor pressure is an important indicator of alkane's volatility, since higher vapor pressure means that an alkane has a higher boiling point at fixed external pressure so will be more stable in an engine. To model vapor pressure as a function of temperature, scientists first record vapor pressure at various temperatures before they fit it to the Antoine equation and determine the coefficients of Antoine equation:

\begin{equation}
\log_{10} p=A-\frac{{B}}{C+T}.
\label{eq:Antoine}
\end{equation}

\noindent Coefficients $A$ and $B$ arise from the solution to the Clausius-Clapeyron relation in an ideal gas approximation, while coefficient $C$ is empirical and captures the temperature dependence of latent heat. Temperature $T$ is measured in $^\circ$C. Experimentally deduced values for $A$,$B$ and $C$ (51,72,72 data entries coming from \cite{TRC}) in our database give an accurate description of vapor pressure's temperature profile between temperatures at which $\log_{10}p=-1.875$ and $\log_{10}p=0.294$, with pressure measured in bars.

\begin{figure}
	\centering
	\includegraphics[width=0.7\linewidth]{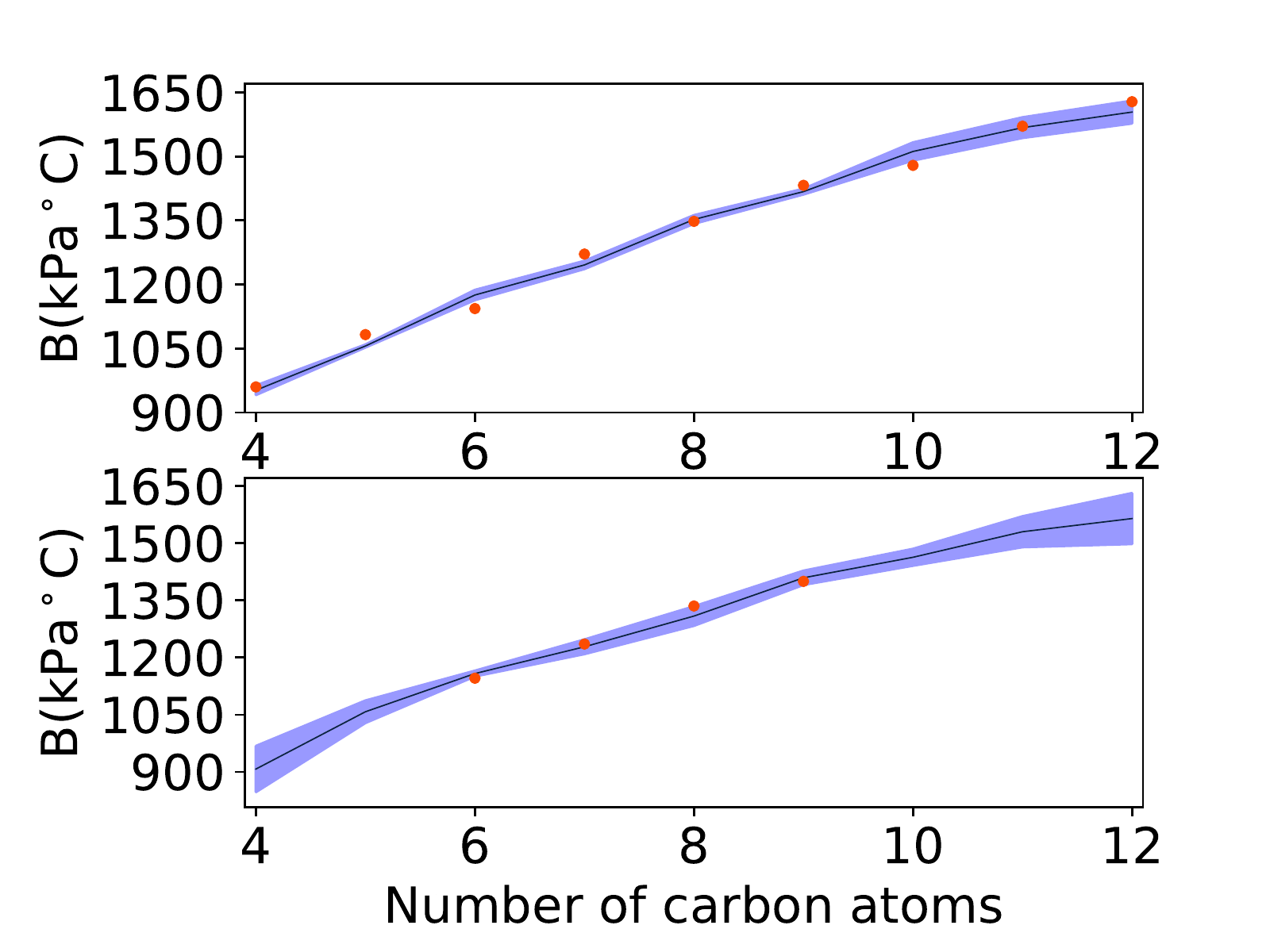}
	\caption{Antoine B coefficient for linear alkanes and 2-methyl homologous series. Dots represent experimental data entries, blue and green lines represent the mean predictions and the colored area represents the uncertainty.}
	\label{fig:AntB}
\end{figure}

To determine the coefficients $B$ and $C$, we train a neural network with 6 hidden nodes for each coefficient. Then, we use the results for $B$,$C$ and the boiling point at atmospheric pressure (when $\log_{10}p=0$) in order to calculate $A$. We use molecular basis as our input nodes and obtain a cross-validation $\textit{R}^2=0.974$ for $B$, $\textit{R}^2=0.962$ for $C$, and $\textit{R}^2=0.958$ for $A$. We also obtain an AAD of 0.008 bar for $A$, 19.81 bar$^{\circ}$C for $B$ and 1.31$^{\circ}$C for $C$. Adding a branch decreases $B$ by about 36 bar$^\circ$C (\autoref{fig:AntB}), adding a branch and keeping molecular weight constant increases $C$ by 3.5$^\circ$C, while extending a branch or moving it along a longest chain is negligible. Antoine $A$ coefficient is approximately constant for all the alkanes, which is consistent with the Clausius-Clapeyron equation, in which $A$ arises as an integration constant. We observe that adding a branch while keeping the molecular weight constant increases the vapor pressure, while extending it and keeping the molecular weight constant or moving it along the longest carbon chain further increases it by a smaller amount than adding a branch.

We use neural network predictions for Antoine coefficients to calculate the vapor pressure as a function of temperature and compare to experimental results. Since vapor pressure is a continuous variable, we use two replacement metrics instead of AAD and  $\textit{R}^2$ to determine the accuracy of our model. To calculate the first metric, we first calculate the following quantity:

\begin{equation}
\gamma=\frac{1}{\Delta T}\int_{T_{\rm min}}^{T_{\rm max}}|p_{\rm exp}(T)-p_{\rm model}(T)|dT,
\end{equation}

before we average it over all the molecules. Note that $\gamma$ is the average absolute deviation of the vapor pressure over the considered temperature range, while $T_{\rm min}$ and $T_{\rm max}$ are calculated via the following relation:

\begin{equation}
(T_{\rm min},T_{\rm max})=\frac{B_{\rm exp}}{A_{\rm exp}-(-1.875,0.294)}+C_{\rm exp}
\end{equation}

Instead of the coefficient of determination, for each molecule we first calculate the following two quantities:

\begin{equation}
\delta^2=\int_{T_{\rm min}}^{T_{\rm max}}(p_{\rm exp}(T)-p_{\rm model}(T))^2dT
\end{equation}

and 

\begin{equation}
\sigma^2=\int_{T_{\rm min}}^{T_{\rm max}}(p_{\rm exp}(T)-\overline{p}_{\rm exp})^2dT,
\end{equation}

where 

\begin{equation}
\overline{p}_{\rm exp}=\frac{1}{\Delta T}\int_{T_{\rm min}}^{T_{\rm max}}p_{\rm exp}(T)dT,
\end{equation}

before we calculate the substitute metric as $1-(\sum_{i}\delta_{i}^2)/(\sum_{i}\sigma_{i}^2)$. This metric tells us how much better our model is compared to a model in which we use an average value of the vapor pressure to model it over an entire temperature range. The value of the former metric is 0.069 bar with a standard deviation of 0.085 bar, while the value of the latter metric is 0.917, indicating a good fit.

\subsection{Flash point}\label{res3}
Flash point is the smallest temperature at which a substance spontaneously ignites in the presence of fire. Predicting it enables us to identify temperatures for lubricant storage and handling.

\begin{figure}
	\centering
	\includegraphics[width=0.7\linewidth]{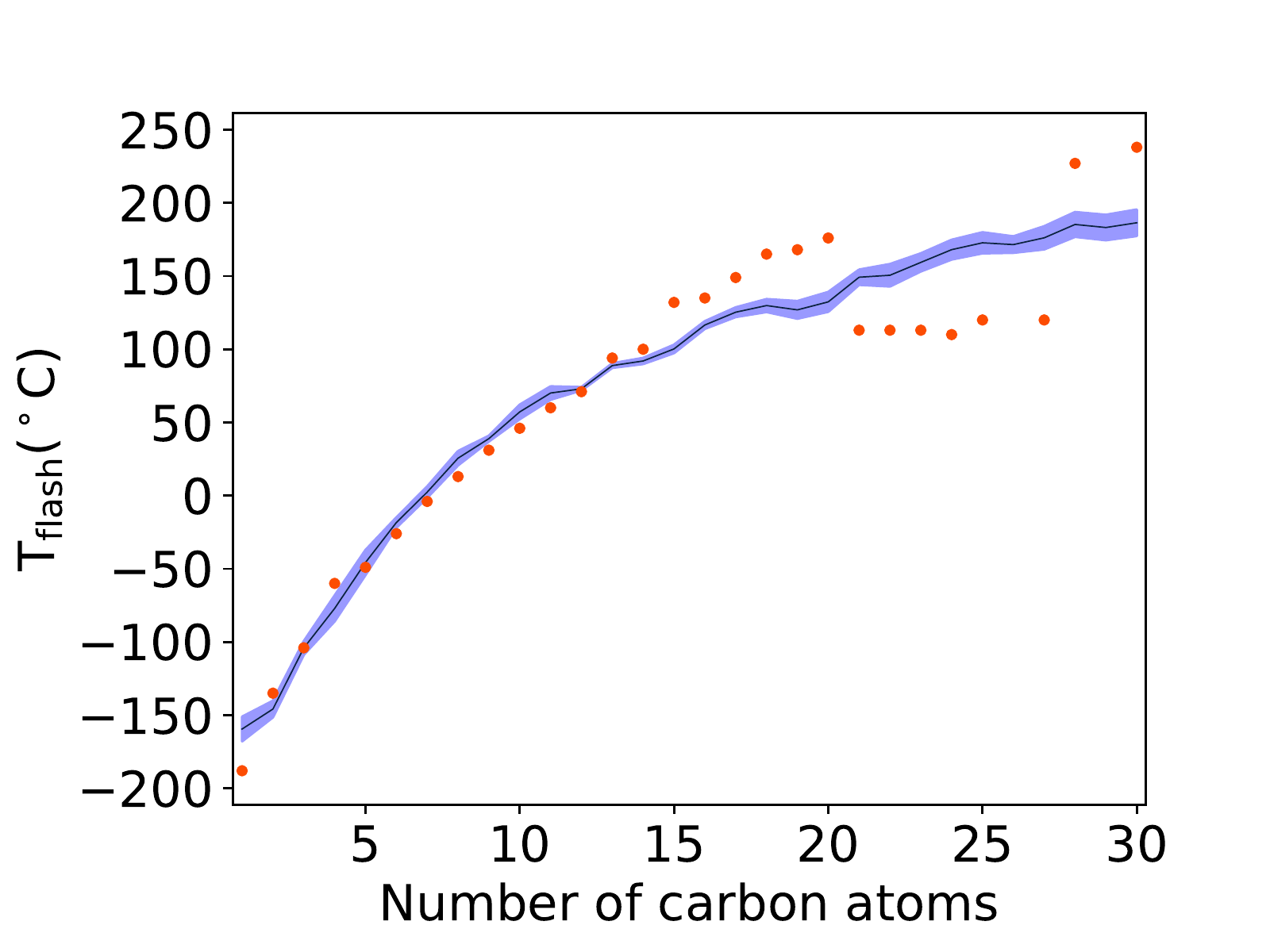}
	\caption{Flash point vs number of carbon atoms when erroneous entries are included. Orange dots represent experimental data points, blue line represents mean prediction, while the colored area around it represents the uncertainty.}
	\label{fig:FlashPoint}
\end{figure}

We study the flash point of linear alkanes with fewer than 31 carbon atoms. We collected experimental data from two online sources \cite{nat}, \cite{chemeo}. After training a neural network with two hidden nodes, we obtain a cross-validation $\textit{R}^2=0.910$. However, we can use neural networks to improve the prediction accuracy. We identify the data entries that lie more than 2 standard errors away from the expected value. In \autoref{fig:FlashPoint}, we see that alkanes that have between twenty and twenty-seven carbon atoms are multiple standard errors away from mean predictions and appear anomalous. After tracking down the original sources of this data \cite{chemeo}, we found that the entries for alkanes from eicosane up to hexacosane are indeed incorrect. We further validate this claim by investigating the correlation between the flash point and the boiling point. It is empirically true that flash point is linearly correlated with the boiling point for hydrocarbon compounds \cite{FPBP}, and linear alkane data entries fit this trend. 

\begin{table*}[ht]
	\begin{center}
		\begin{tabular}{|c|c|c|c|c|c|} 
			\hline
			\textbf{Molecule} & \textbf{\makecell{Experimental \\ ($^{\circ}$C)}} & \textbf{\makecell[c]{Group Contribution \\ Method \cite{Mathieu}  \\ Prediction ($^{\circ}$C)}} & \textbf{\makecell[c]{Neural Network \\ Prediction ($^{\circ}$C)}} & \textbf{\makecell[c]{Group \\ Contribution \\ Method \cite{Mathieu} \\ Absolute \\ Deviation ($^{\circ}$C)}} & \textbf{\makecell[c]{Neural Network \\ Absolute \\ Deviation ($^{\circ}$C)}}\\
			\hline
			Ethane	&	-139.16	&	-129.04	&	-137.63	&	10.12	&	1.53	\\
			Propane	&	-106.49	&	-97.15	&	-106.36	&	9.34	&	0.13	\\
			Butane	&	-74.00	&	-71.15	&	-73.47	&	2.85	&	0.53	\\
			Pentane	&	-47.21	&	-47.15	&	-46.23	&	0.06	&	0.98	\\
			Hexane	&	-17.40	&	-26.15	&	-23.02	&	8.75	&	5.62	\\
			Heptane	&	-7.12	&	-6.15	&	-2.60	&	0.97	&	4.52	\\
			Octane	&	16.15	&	11.85	&	16.00	&	4.30	&	0.15	\\
			Nonane	&	29.29	&	28.85	&	33.52	&	0.44	&	4.23	\\
			Decane	&	50.45	&	44.85	&	50.61	&	5.60	&	0.16	\\
			Undecane	&	69.45	&	60.85	&	67.89	&	8.60	&	1.56	\\
			Dodecane	&	85.43	&	74.85	&	83.92	&	10.58	&	1.51	\\
			Tridecane	&	100.32	&	89.85	&	98.21	&	10.47	&	2.11	\\
			Tetradecane	&	111.30	&	102.85	&	111.24	&	8.45	&	0.06	\\
			Pentadecane	&	122.55	&	115.85	&	123.34	&	6.70	&	0.79	\\
			Hexadecane	&	131.67	&	128.85	&	134.65	&	2.82	&	2.98	\\
			Heptadecane	&	146.83	&	141.18	&	145.41	&	5.65	&	1.42	\\
			Octadecane	&	156.28	&	153.14	&	155.81	&	3.14	&	0.47	\\
			Nonadecane	&	167.25	&	164.79	&	165.80	&	2.46	&	1.45	\\
			Icosane	&	175.96	&	176.13	&	175.34	&	0.17	&	0.62	\\
			Octacosane	&	226.81	&	258.19	&	228.83	&	31.38	&	2.01	\\
			Triacontane	&	239.93	&	276.80	&	238.23	&	36.87	&	1.70	\\
			\hline
		\end{tabular}
		\caption{Experimental values and prediction of the flash point  for indicated molecules. The table compares the accuracy of our neural network model with the accuracy of a model based on the group contribution method.}
		\label{tab:table3}
	\end{center}
\end{table*}

\begin{figure}
	\centering
	\includegraphics[width=0.7\linewidth]{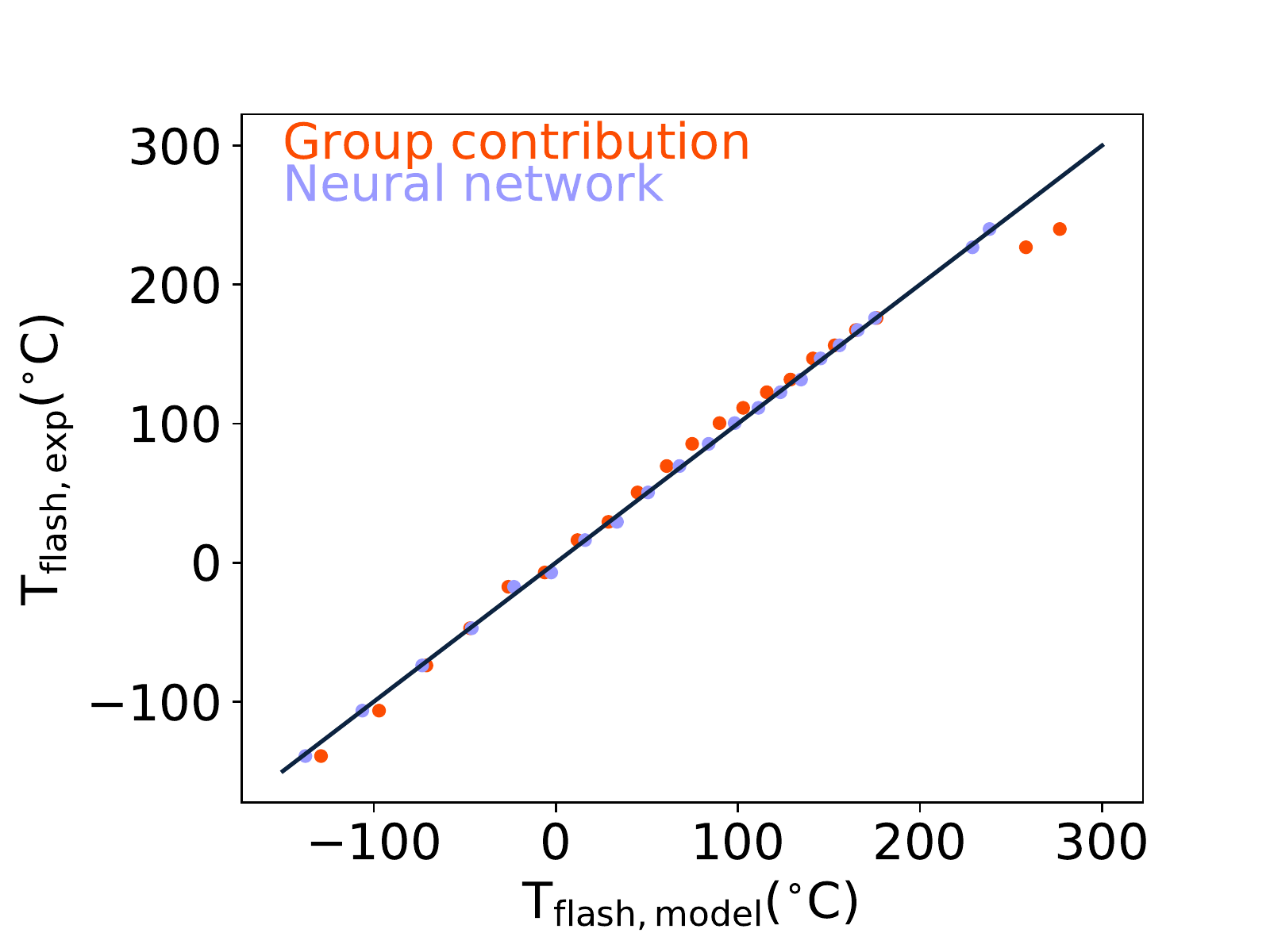}
	\caption{Parity plot for the flash point. Our neural network model is compared to a model based on the group contribution method \cite{Mathieu}.}
	\label{fig:FPPar}
\end{figure}

After removing erroneous data entries, we predict flash point again and obtain a cross validation $\textit{R}^2=0.999$, which would allow the model predictions to replace experimental measurements. We also compare our predictions to those made by a group contribution method presented in \cite{Mathieu}. The neural network model reproduces experimental flash point with an AAD of 1.65$\rm{^{\circ}C}$, compared to an AAD of 8.08$\rm{^{\circ}C}$ predicted by a group contribution method (\autoref{tab:table3}). Our model gives a more accurate prediction for 16 out of 21 alkanes used to build a model. In addition, our model shows greater consistency than the group contribution model. In particular, the model in \cite{Mathieu} is far less accurate for the several smaller molecules such as ethane and propane, as well as octacosane and triancontane, whose flash point is mispredicted by over 30$^{\circ}$C, while the accuracy of our model roughly consistent for all the data entries (\autoref{fig:FPPar}).

\subsection{Melting point} \label{res4}
Accurate predictions of the melting point reveals whether an alkane will solidify at lower temperatures of the lubricant's operating range. We study the melting point of branched alkanes with fewer than 13 carbon atoms and train a neural network with 5 hidden nodes and just the molecular basis as input nodes. Our dataset consists comprises 51 molecules, whose melting point was experimentally measured \cite{TRC}. After training a neural network model and cross-validating it we obtain an $\textit{R}^2=0.650$. This poor reproduction of experimental data motivates us to search for additional physical correlations to improve the fidelity of the neural network predictions.

We have identified two additional effects that affect the melting point. Firstly, if the number of carbon atoms in the longest carbon chain is an even number, an alkane has a higher melting point than if the number of carbon atoms is odd. Secondly, an alkane with a higher number of molecular symmetries has a higher melting point. This effect is readily observed in isomers of pentane \cite{MolSim}. Pentane has 4 molecular symmetries and a melting point of -129.9$^\circ$C, 2-methylbutane has 2 molecular symmetries and a a melting point of -159$^\circ$C, while 2,2-dimethylpropane has 24 molecular symmetries and a melting point of -16.6$^\circ$C. Therefore, we add two more elements to the input layer of the neural network that we use to predict the melting point; one to capture the odd/even effect, and the second being the total number of symmetries.

\begin{figure}
	\centering
	\includegraphics[width=0.7\linewidth]{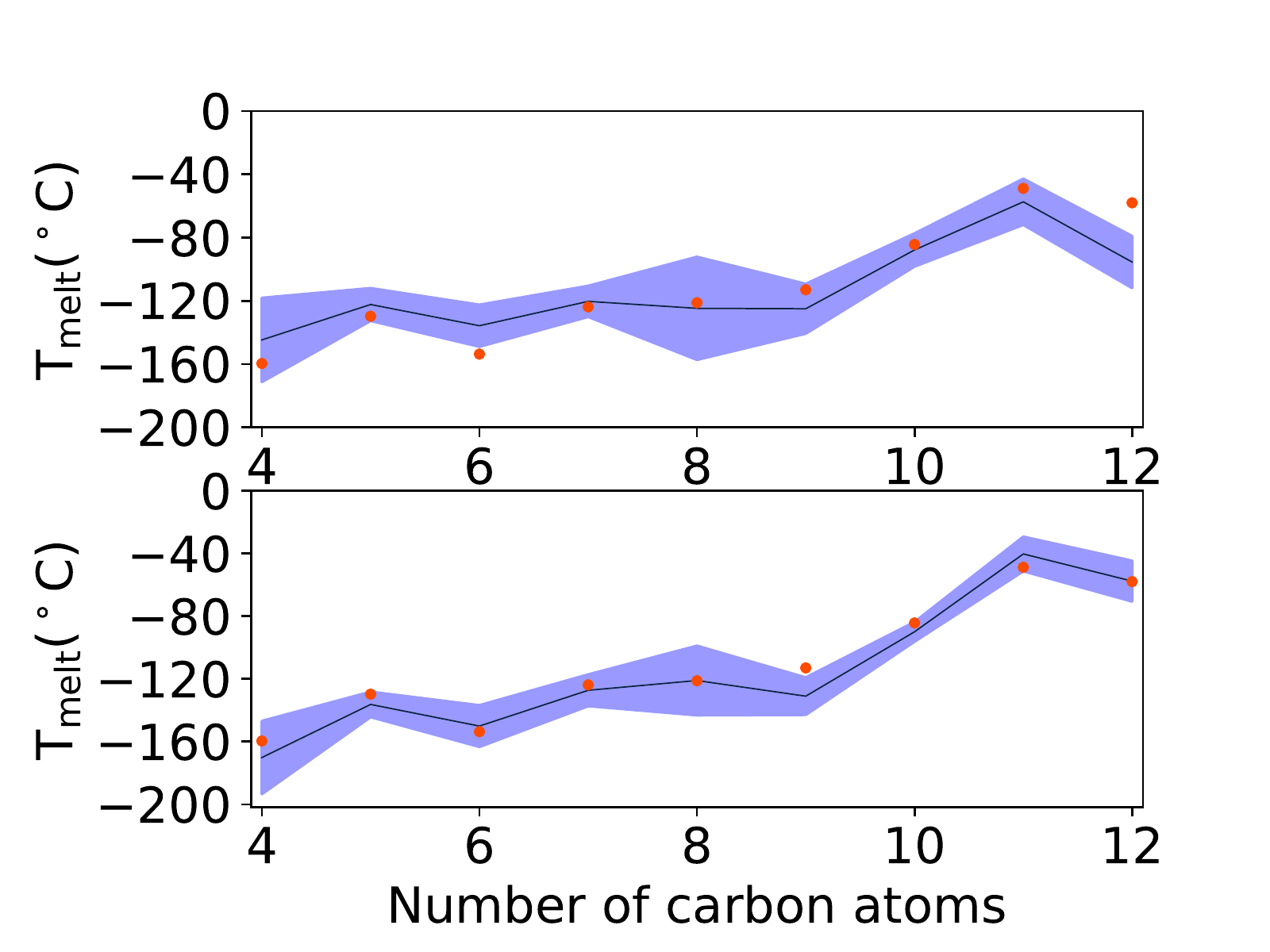}
	\caption{Melting point prediction for randomly selected alkanes. In the top figure, we show predictions obtained from a neural network model that used only molecular basis as input nodes. In the bottom figure, we show predictions obtained from a neural network model when two symmetry-based input nodes have been added. Orange dots show experimental data, blue line shows the mean prediction while colored area shows the uncertainty region.}
	\label{fig:MP}
\end{figure}

With these two additional chemical descriptors, we train and cross-validate a new neural network with a cross-validation, obtaining $\textit{R}^2=0.998$ (\autoref{fig:MP}). We also compare our results to results obtained by several regression models that use molecular structure and topological indices as inputs \cite{MeltingPoint} for 4 molecules whose melting point is common to both datasets (\autoref{tab:table4}). Neural network model shows greater accuracy, as it reproduces experimental values with an AAD of 1.45$^{\circ}$, compared to an AAD of 3.45$^{\circ}$ reproduced by models 4.1 and 4.2 in \cite{MeltingPoint}.

\begin{table*}[ht]
	\begin{center}
		\begin{tabular}{|c|c|c|c|c|c|} 
			\hline
			\textbf{Molecule} & \textbf{\makecell{Experimental \\ ($^{\circ}$C)}} & \textbf{\makecell[c]{Topological \\ Indices \cite{ViscosityComparison} \\ Prediction ($^{\circ}$C)}} & \textbf{\makecell[c]{Neural Network \\ Prediction ($^{\circ}$C)}} & \textbf{\makecell[c]{Topological \\ Indices \cite{ViscosityComparison} \\ Absolute \\ Deviation ($^{\circ}$C)}} & \textbf{\makecell[c]{Neural Network \\ Absolute \\ Deviation ($^{\circ}$C)}}\\
			\hline
			4-methylnonane	&	-98.70	&	-95.15	&	-95.66	&	3.55	&	3.04	\\
			Dodecane	&	-9.58	&	-11.25	&	-10.44	&	1.67	&	0.86	\\
			2-methylundecane	&	-46.81	&	-47.85	&	-46.17	&	1.04	&	0.64	\\
			3-methylundecane	&	-58.00	&	-65.55	&	-56.74	&	7.55	&	1.26	\\
			
			\hline
		\end{tabular}
		\caption{Experimental values and prediction of the melting point  for indicated molecules. The table compares the accuracy of our neural network model with the accuracy of regression models based on the topological indices and molecular structure.}
		\label{tab:table4}
	\end{center}
\end{table*}

The significant improvement in accuracy of our model upon the introduction of the number of symmetries serves as a further indicator of the importance of the molecular symmetry on the melting point of alkanes. Looking forward, to further improve the accuracy of the predictions, one would also include the details of alkanes' crystalline structure.

\subsection{Kinematic viscosity}\label{res5}
Dynamic viscosity ($\mu$) is a measure of a fluid's resistance to an external force. The ratio of dynamic viscosity and density ($\rho$) gives the kinematic viscosity, a measure of fluid's flow properties. Predicting an alkane's kinematic viscosity at 40$^\circ$ and 100$^\circ$C enables us to calculate its viscosity index (VI)\cite{VI}, frequently used in industry as a measure of temperature gradient of kinematic viscosity.

We study the kinematic viscosity of the linear alkanes from heptane to heneicosane. We first train a neural network to predict dynamic viscosity and density using an experimental database assembled by merging experimental data for shear viscosity and density as a function of temperature and pressure obtained from various research papers
(\cite{Assael1991}, \cite{BALED2014108}, \cite{Caudwell2004}, \cite{doi:10.1021/je800417q}, \cite{HERNANDEZGALVAN200751}, \cite{doi:10.1021/je00049a011}, \cite{SANTOS201746}, \cite{unknown}) into a single dataset. Our dataset for density has 537 data entries while our dataset for viscosity has 638 data entries. We then take the ratio of the predictions to determine the kinematic viscosity and its uncertainty.

\begin{figure}
	\centering
	\includegraphics[width=0.7\linewidth]{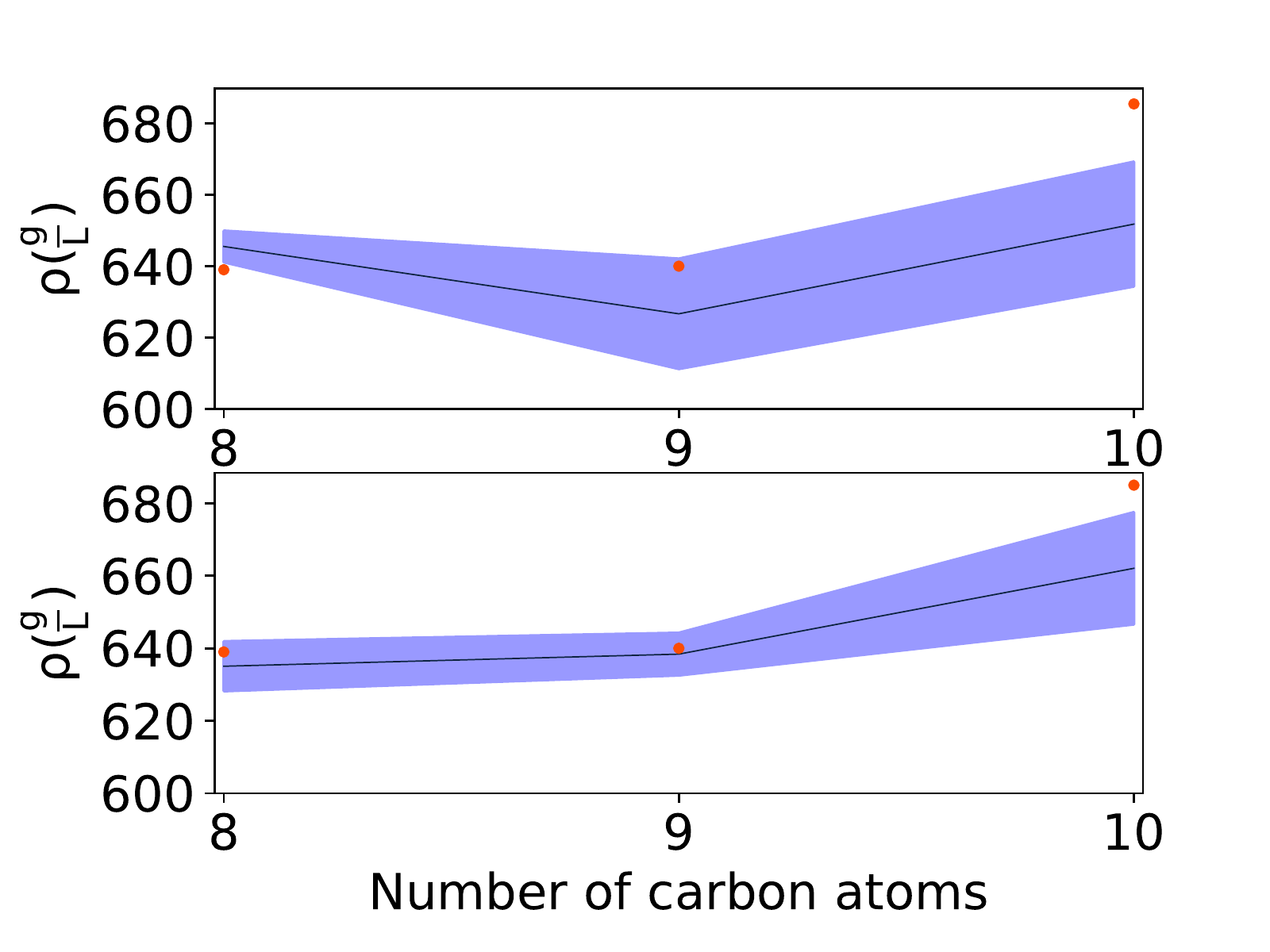}
	\caption{Cross-validation of density at 100$^\circ$C for 4-methylheptane, 2,6-dimethylheptane and 4,5-dimethyloctane. In the top figure, we show results obtained from the neural network architecture that uses only the molecular basis as input features. In the bottom figure, we show results obtained from the neural network architecture that uses molecular basis and density data at 20$^\circ$C as input features. }
	\label{fig:DynamicViscosityAndDensityS}
\end{figure}

Data for dynamic viscosity and density as a function of temperature and pressure is fragmented, so we need more input parameters for the neural network models. To predict density and dynamic viscosity at 40$^\circ$C and 100$^\circ$C, we include the additional sparse data of density and dynamic viscosity at 25$^\circ$C and atmospheric pressure. For the branched alkanes with eight, nine and ten carbon atoms, the neural network model for density as a function of temperature  increased in accuracy from $\textit{R}^2=0.412$ to $\textit{R}^2=0.840$ due to the inclusion of this additional information(\autoref{fig:DynamicViscosityAndDensityS}). Furthermore, focusing just on linear alkanes we obtain a cross-validation $\textit{R}^2$ of 0.998 for dynamic viscosity at 20\degree C and of 0.987 for density. 

\begin{figure}
	\centering
	\includegraphics[width=0.7\linewidth]{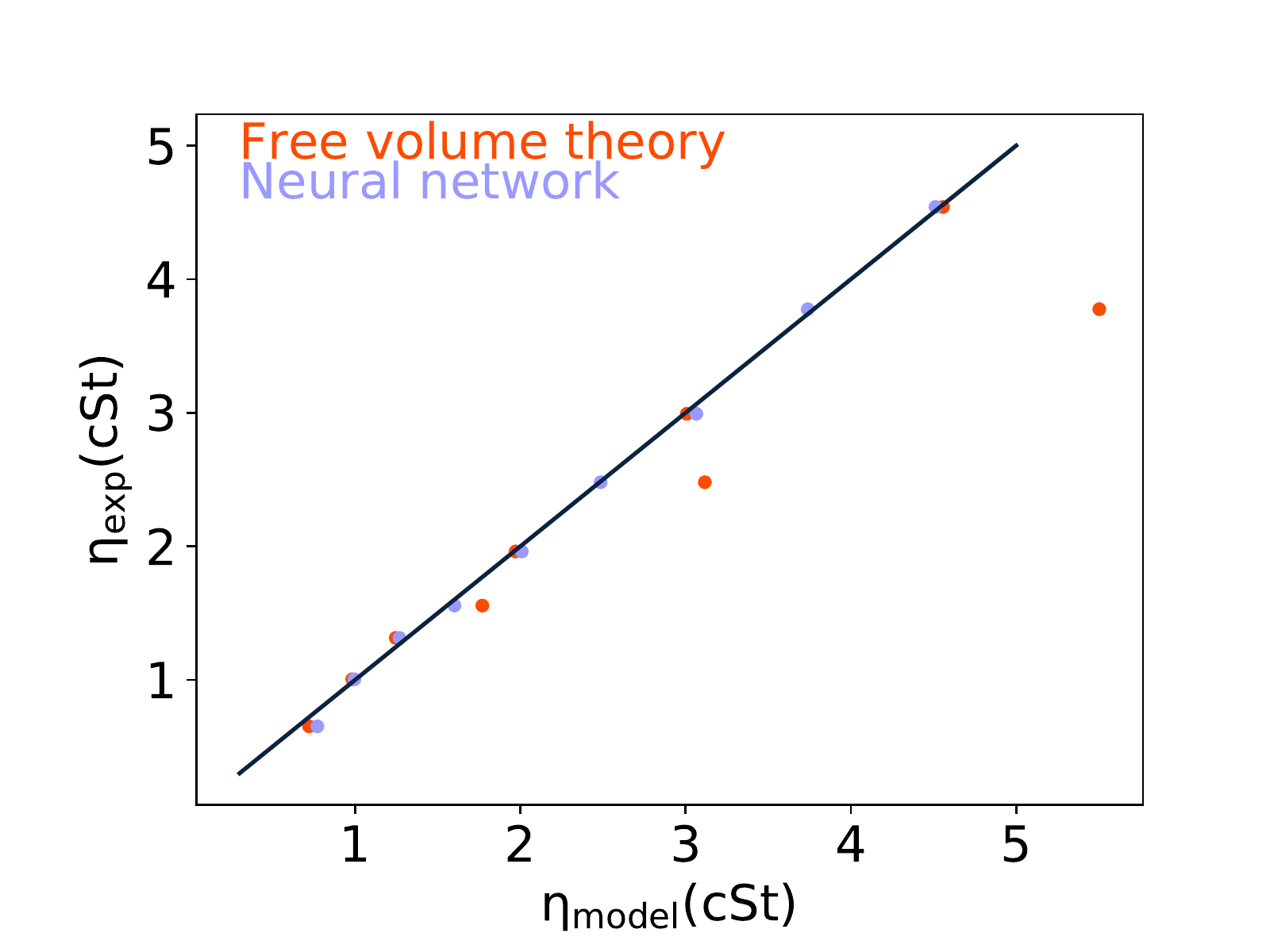}
	\caption{Parity plot for kinematic viscosity of linear alkanes. Our neural network model is compared to a model based on free volume theory \cite{ViscosityComparison}.}
	\label{fig:ViscPar}
\end{figure}

Next, we compare the values for kinematic viscosity at 20\degree C and atmospheric pressure obtained by our ANN to values obtained by a model based on free volume theory \cite{ViscosityComparison} (\autoref{tab:table5}). The neural network model is more accurate than the free volume theory model, reproducing experimental data with $\textit{R}^2=0.998$ and an average absolute deviation of 0.05 cSt, compared to $\textit{R}^2=0.899$ and an average absolute deviation of 0.31 cSt predicted by the free volume theory model \cite{ViscosityComparison}. Furthermore, the neural network model shows greater consistency for the molecules analysed than the free volume theory model. Standard deviation in absolute deviation of the neural network model is 0.03 cSt compared to 0.57 cSt for the free volume theory model, with absolute deviations in kinematic viscosity of pentadecane(1.73 cSt) and tridecane(0.64 cSt) being particularly large (\autoref{fig:ViscPar}).

\begin{figure}
	\centering
	\includegraphics[width=0.7\linewidth]{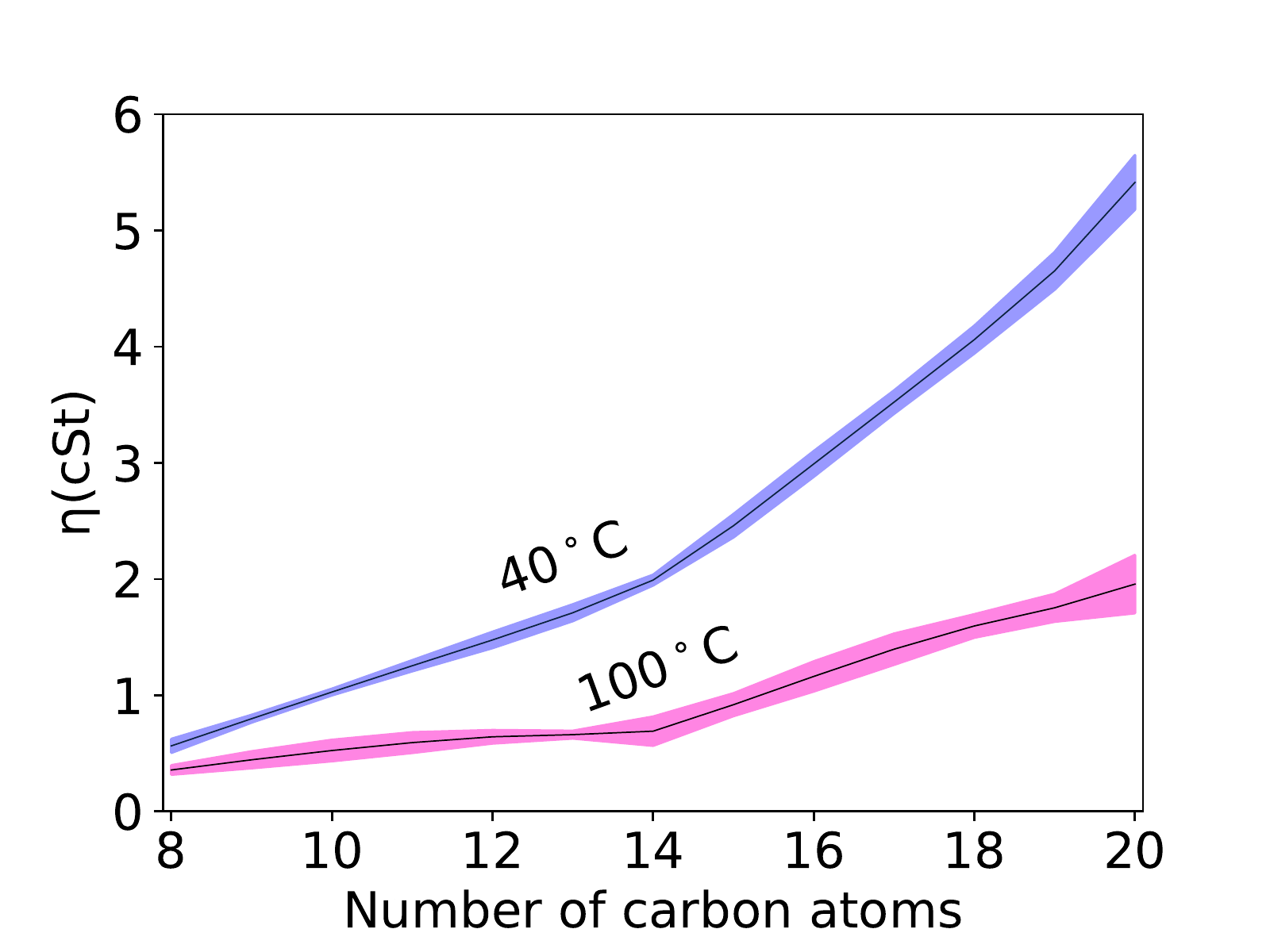}
	\caption{Blue/cyan plot shows predictions for the kinematic viscosity at 40\degree C while black/magenta shows predictions at 100\degree C.}
	\label{fig:KinematicViscosityS}
\end{figure}

\begin{table*}[ht]
	\begin{center}
		\begin{tabular}{|c|c|c|c|c|c|} 
			\hline
			\textbf{Molecule} & \textbf{\makecell{Experimental \\ (cSt)}} & \textbf{\makecell[c]{Free Volume \\ Theory Model \cite{ViscosityComparison}  \\ Prediction (cSt)}} & \textbf{\makecell[c]{Neural Network \\ Prediction (cSt)}} & \textbf{\makecell[c]{Free Volume \\ Theory Model \cite{ViscosityComparison} \\ Absolute \\ Deviation (cSt)}} & \textbf{\makecell[c]{Neural Network \\ Absolute \\ Deviation (cSt)}}\\
			\hline
			Octane	&	0.65	&	0.72	&	0.77	&	0.07	&	0.12	\\
			Nonane	&	1.00	&	0.98	&	0.99	&	0.02	&	0.01	\\
			Decane	&	1.31	&	1.25	&	1.27	&	0.07	&	0.05	\\
			Undecane	&	1.56	&	1.77	&	1.60	&	0.21	&	0.05	\\
			Dodecane	&	1.96	&	1.97	&	2.01	&	0.01	&	0.05	\\
			Tridecane	&	2.48	&	3.12	&	2.49	&	0.64	&	0.01	\\
			Tetradecane	&	2.99	&	3.01	&	3.06	&	0.01	&	0.07	\\
			Pentadecane	&	3.78	&	5.50	&	3.74	&	1.73	&	0.04	\\
			Hexadecane	&	4.54	&	4.56	&	4.51	&	0.02	&	0.03	\\

			\hline
		\end{tabular}
		\caption{Experimental values and prediction of the kinematic viscosity at 20 \degree C and atmospheric pressure for indicated molecules. The table compares the accuracy of our neural network model with the accuracy of a model based on the free volume theory.}
		\label{tab:table5}
	\end{center}
\end{table*} 

Finally, we run our neural network model on density and dynamic viscosity at 40$^\circ$C and 100$^\circ$C and at atmospheric pressure to calculate the kinematic viscosity at 40$^\circ$C and 100$^\circ$C (\autoref{fig:KinematicViscosityS}) at atmospheric pressure and then determine alkane's viscosity index. The neural network model can provide insights into which linear alkanes could feature in a commercialized lubricant. Eicosane is the only linear alkane modelled here that has a value of kinematic viscosity at 100$^\circ$C above 2 $\rm{cSt}$ so it is the only linear alkane for which we can define a viscosity index. However, eicosane is a solid below 36$^\circ$C so could only be present in a base oil lubricant in relatively small amounts, as lubricants are usually expected to operate between -15$^\circ$C and 100$^\circ$C. Therefore, it is likely that linear alkanes are present in base oil lubricants only in relatively small amounts.

\section{Conclusions} \label{Conclusions}
We have used artificial neural networks that exploit inter-property correlations to predict the physical properties of alkanes. The algorithm describes the molecular structure of linear, single, and double branched alkanes, and enables us to predict the boiling point, the heat capacity and the vapor pressure as a function of temperature. We also predicted the flash point of linear alkanes up to tridecane and identified erroneous experimental entries in the literature. The number of molecular symmetries correlates to the melting point. Finally, we have exploited the temperature and pressure dependence of dynamic viscosity and density alongside interproperty correlations across the temperature range to predict the kinematic viscosity at atmospheric pressure as a function of temperature. Values of physical properties reproduced with these neural networks are more accurate and consistent than the values reproduced by other methods. We present a summary of our results for the boiling point, the molar heat capacity, the Antoine coefficients, the flash point, the melting point and the kinematic viscosity in \autoref{tab:tableX}.

\begin{table*}[!h]
	\begin{center}
		\begin{tabular}{|c|c|c|c|} 
			\hline
			\textbf{\makecell[c]{Physical \\ Property}} & {$\rm{N_{molecules}}$} & \textbf{$\textit{R}^2$} & \textbf{AAD} \\
			\hline
			$\rm{T_{boil}}$ 	&	188	&	0.992	&	 1.74 $\rm{^{\circ}}$C \\
			$\rm{C_{molar}}$ 	&	181	&	0.997	&	 2.33 $\rm{J(molK)^{-1}}$ \\
			$\rm{T_{flash}}$ 	&	21	&	0.999	&	 1.61 $^{\circ}$C \\
			$\rm{T_{melting}}$ 	&	51	&	0.998	&	 1.26 $^{\circ}$C \\
			$\rm{p_{vapor}} $ 	&	51	&	0.917	&	 0.07 bar\\
			$\rm{\nu}$ 	&	9	&	0.998	&	 0.05 cSt \\
			
			\hline
		\end{tabular}
		\caption{Summary of results for all the physical properties analysed.}
		\label{tab:tableX}
	\end{center}
\end{table*} 

Our study serves as a solid platform from which to further investigate physical properties of alkanes. This generic neural network architecture could merge sparse experimental data with molecular dynamics simulations to predict physical properties of alkanes, particularly the intractable properties like shear viscosity and density, enabling us to identify the alkanes that could be components for lubricant base oils with superior physical properties.

Pavao Santak acknowledges financial support of BP-ICAM. Gareth Conduit acknowledges financial support from the Royal Society. Both authors thank Leslie Bolton, Corneliu Buda and Nikolaos Diamantonis for useful discussions. There is an Open Access at \url{https://www.openaccess.cam.ac.uk}.

\nocite{*}
\bibliographystyle{ieeetr}
\bibliography{bibliography}

\end{document}